\documentclass[12pt]{iopart}

\usepackage{iopams}
\usepackage{graphicx}
\usepackage{subfig}
\usepackage{color}

\begin{document}

\title{Low-$n$ global ideal MHD instabilities in CFETR baseline scenario}

\author{Rui HAN}
\address{CAS Key Laboratory of Geospace Environment and Department of Engineering and Applied Physics, University of Science and Technology of China, Hefei 230026, China}

\author{Ping ZHU}
\address{International Joint Research Laboratory of Magnetic Confinement Fusion and Plasma Physics, State Key Laboratory of Advanced Electromagnetic Engineering and Technology, School of Electrical and Electronic Engineering, Huazhong University of Science and Technology, Wuhan, Hubei 430074, China}
\address{Department of Engineering Physics, University of Wisconsin-Madison, Madison, Wisconsin 53706, USA}
\ead{zhup@hust.edu.cn}

\author{Debabrata BANERJEE}
\address{CAS Key Laboratory of Geospace Environment and Department of Engineering and Applied Physics, University of Science and Technology of China, Hefei 230026, China}

\author{Shikui CHENG}
\address{CAS Key Laboratory of Geospace Environment and Department of Engineering and Applied Physics, University of Science and Technology of China, Hefei 230026, China}

\author{Xingting YAN}
\address{CAS Key Laboratory of Geospace Environment and Department of Engineering and Applied Physics, University of Science and Technology of China, Hefei 230026, China}

\author{Linjin ZHENG}%
\address{Institute of Fusion Studies, University of Texas at Austin, Austin, Texas 78712, USA}

\author{The CFETR Physics Team}


\newpage
\begin{abstract}
This article reports an evaluation on the linear ideal magnetohydrodynamic (MHD) stability of the China Fusion Engineering Test Reactor (CFETR) baseline scenario for various first-wall locations. The initial-value code NIMROD and eigen-value code AEGIS are employed in this analysis. A good agreement is achieved between two codes in the growth rates of $n=1-10$ ideal MHD modes for various locations of the perfect conducting first-wall. The higher-$n$ modes are dominated by ballooning modes and localized in the pedestal region, while the lower-$n$ modes have more prominent external kink components and broader mode profiles. The influences of plasma-vacuum profile and wall shape are also examined using NIMROD. In presence of resistive wall, the low-$n$ ideal MHD instabilities are further studied using AEGIS. For the designed first-wall location, the $n = 1$ resistive wall mode (RWM) is found unstable, which could be fully stabilized by uniform toroidal rotation above 2.9\% core Alfv\'en speed.
\end{abstract}

%
%
%
%
%

\section{Introduction}
\label{sec1}


Besides being a partner in International Thermonuclear Experimental Reactor (ITER)~\cite{aymer2002}, China has recently proposed to design and potentially build China Fusion Engineering Test Reactor~(CFETR)~\cite{wan2014}. The goal is to address the physics and engineering issues essential for bridging the gap between ITER and DEMO (DEMOnstration Power Station), including achieving tritium breeding ratio (TBR) $>1$ and exploring options for DEMO blanket and divertor solutions~\cite{chan2015,shi2016}. A conceptual engineering design of CFETR including different coils and remote maintenance systems was prepared in the beginning~\cite{song2014}. The initial design parameters of CFETR are based on a 0-D analysis~\cite{wan2014}, and later are optimized using several 1.5D transport codes~\cite{chan2015}. To achieve staged goals, the CFETR has been designed for two steady-state scenarios - baseline and advanced scenarios~\cite{shi2016}. The baseline scenario is designed for moderate fusion power (200MW) with a fully non-inductive current drive, giving more importance towards challenging annual duty factor of $0.3-0.5$. The advanced design is aimed at higher fusion power with a substantial challenging fraction of bootstrap current drive.


In the baseline scenario, the current drive sources are deposited far off-axis, and as a result there is a reversed magnetic shear with the minimum of safety factor $q_{min} >2$ located at an outer radius~\cite{jiale2017}. Fully non-inductive operation requires at least $36\%$ of bootstrap current fraction (see Table-1 of~\cite{jian2017}), which leads to high pedestal pressure gradient and peaked edge current. Such a configuration is expected to be unstable to both ballooning and external kink modes~\cite{peeters2000}. In principle, the external kink modes, which have the potential to lead to plasma disruptions, are dominantly low-$n$ modes and may have strong interactions with the first-wall ~\cite{chu2010,igochine2012}. The high-$n$ modes are more dominated by the peeling-ballooning modes that are localized near edge. The edge-localized modes (ELMs) are less dangerous, but the repetitive expulsion of stored plasma energy and particles due to ELMs, would degrade plasma confinement and damage divertor and first-wall components.


To provide physics base for the engineering design on the optimal choice of first-wall position of CFETR, a thorough evaluation of $n=1-10$ modes is performed in this paper. Assuming the plasma is surrounded by a conformal and perfect conducting first-wall, the dependence of growth rates for $n=1-10$ modes on the wall position is evaluated using the initial value extended-MHD code NIMROD~\cite{sovinec2004} and the eigenvalue code AEGIS~\cite{zheng2006}. With a perfect conducting wall located at the designed wall position, the most dangerous $n=1$ mode is found stable. Considering in practice the first-wall is not perfectly conducting, the $n=1$ ideal MHD mode would be actually the RWM, which grows in the wall-resisitive time scale. In last two decades, toroidal plasma rotation has been found to have stabilizing effects om RWMs~\cite{bondeson1994,strait2003,zheng2005,liuyueqiang2013}. We employ AEGIS code to evaluate the rotational stabilizing effects on RWMs in CFETR baseline scenario. The rotation threshold where the $n=1$ RWM growth can be fully suppressed is found.


This paper is organized as follows. In Sec.~\ref{sec2}, the equilibrium profiles of the baseline scenario are introduced. Sec.~\ref{sec3} describes the resistive single-fluid MHD model in NIMROD and AEGIS codes. Sec.~\ref{sec4} reports the calculations for low-$n$ ideal MHD modes in presence of a perfect conducting wall using both NIMROD and AEGIS codes. Effects of more realistic plasma-vacuum profile and wall shape are discussed in Sec.~\ref{sec5}. In Sec.~\ref{sec6}, rotational effects towards the stabilization of RWMs are presented. Conclusions and discussions are given in Sec.~\ref{sec7}.


\section{Equilibrium of CFETR baseline scenario} 
\label{sec2}

\begin{figure}[htbp] 
\centering
\begin{minipage}{0.49\textwidth}
\includegraphics[width=1.0\textwidth]{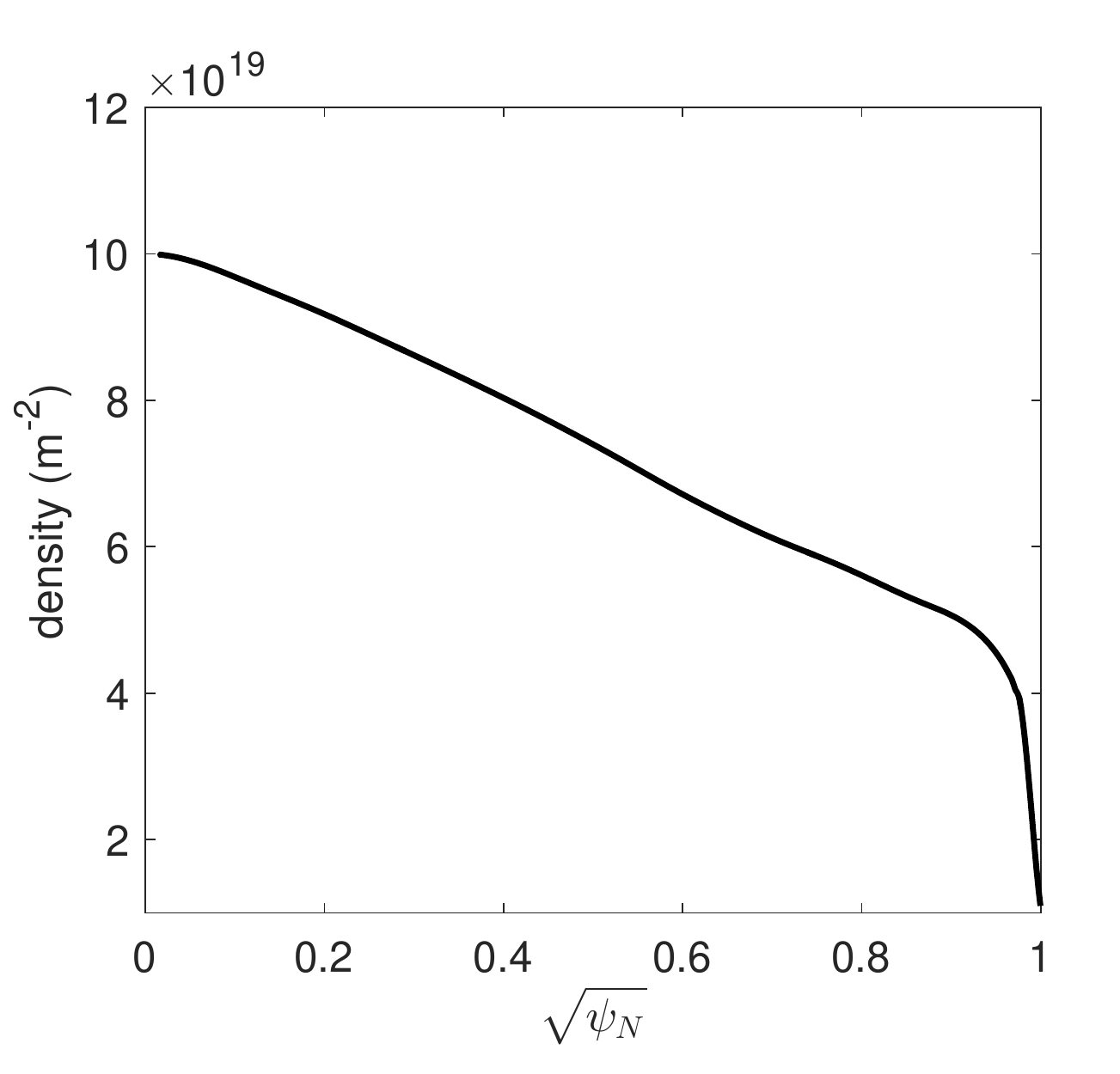}
\put(-220,180){\textbf{(a)}}
\end{minipage}
\begin{minipage}{0.49\textwidth}
\includegraphics[width=1.0\textwidth]{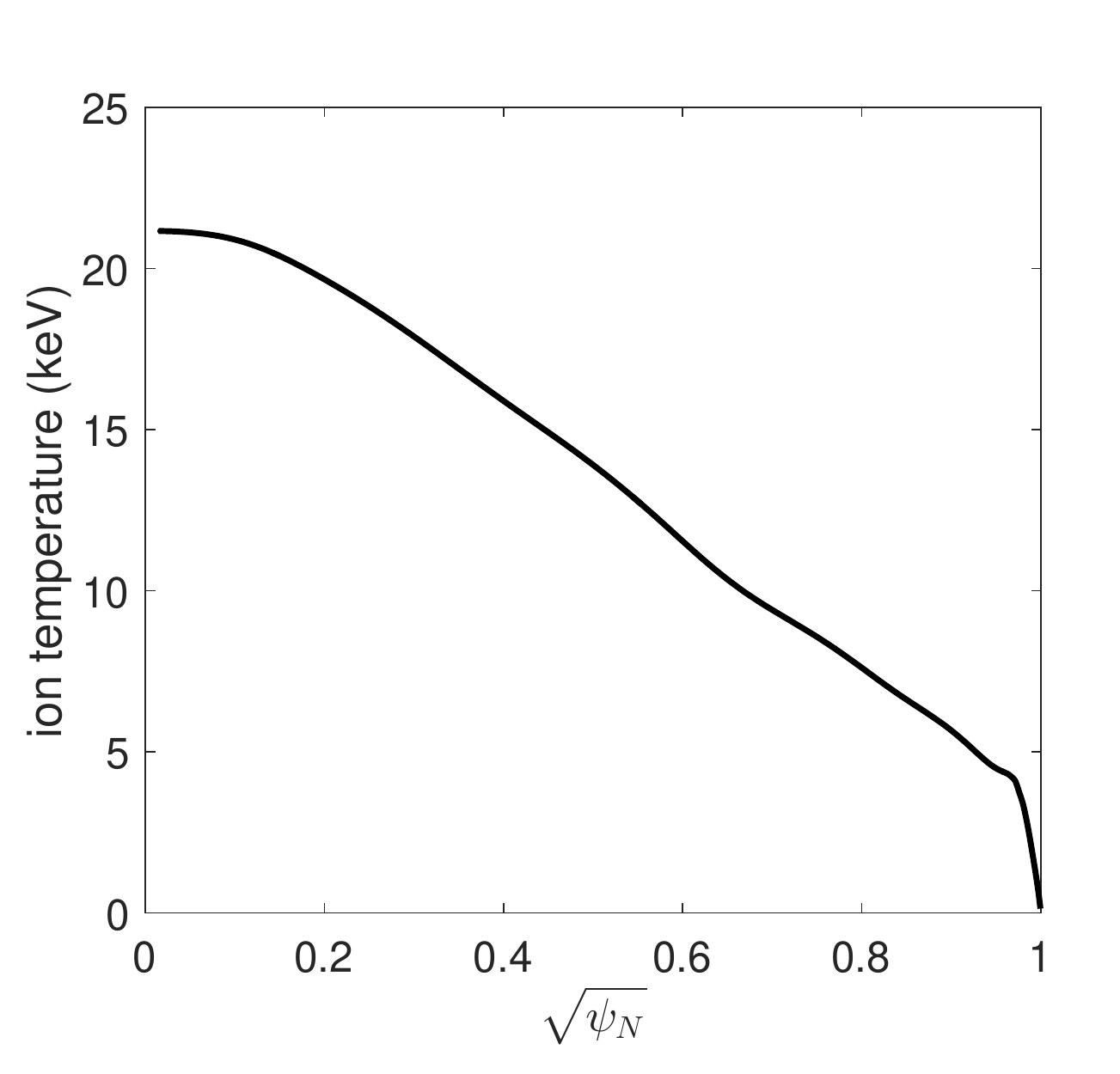}
\put(-220,180){\textbf{(b)}}
\end{minipage}

\begin{minipage}{0.49\textwidth}
\includegraphics[width=1.0\textwidth]{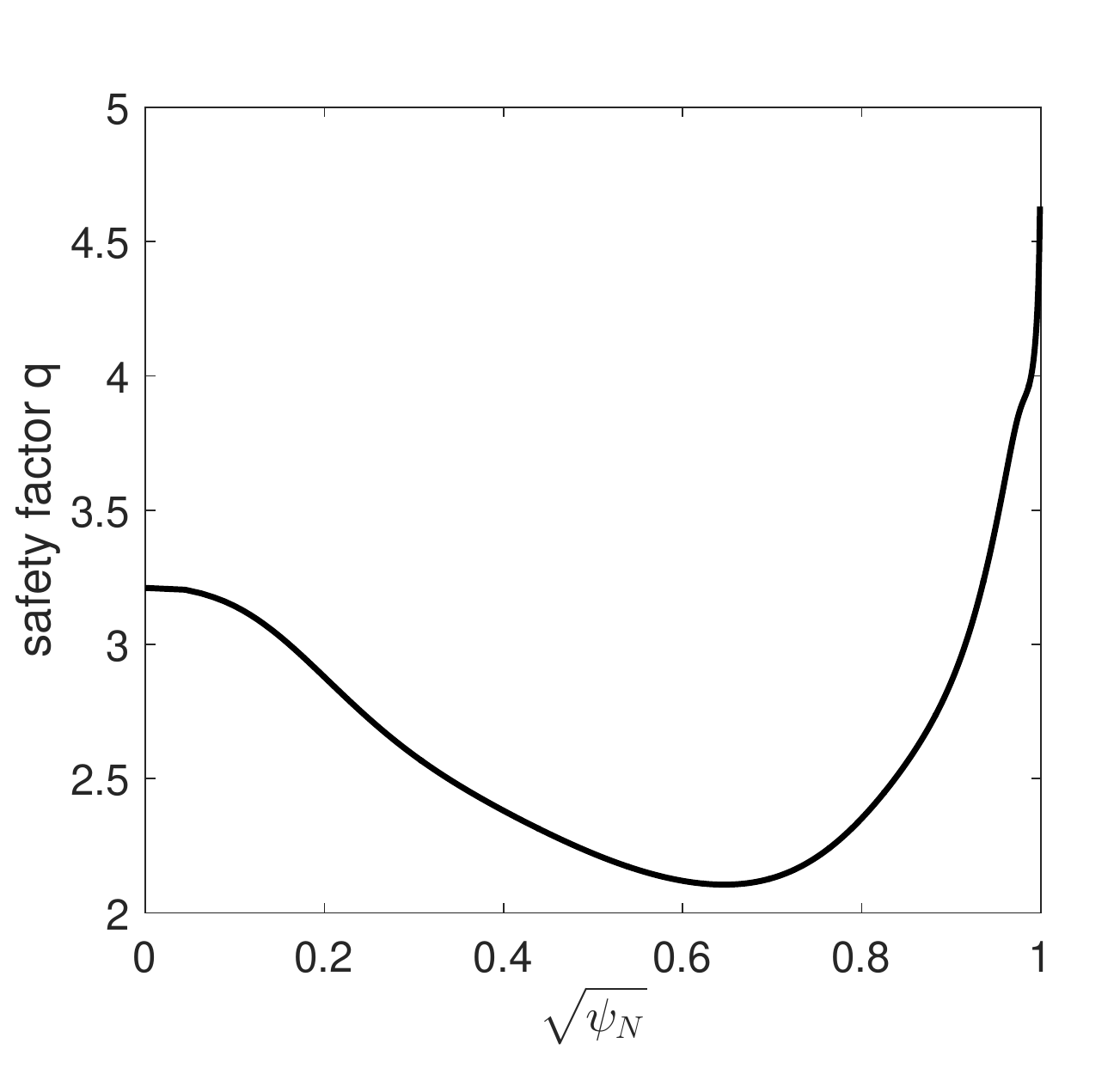}
\put(-220,180){\textbf{(c)}}
\end{minipage}
\begin{minipage}{0.49\textwidth}
\includegraphics[width=1.0\textwidth]{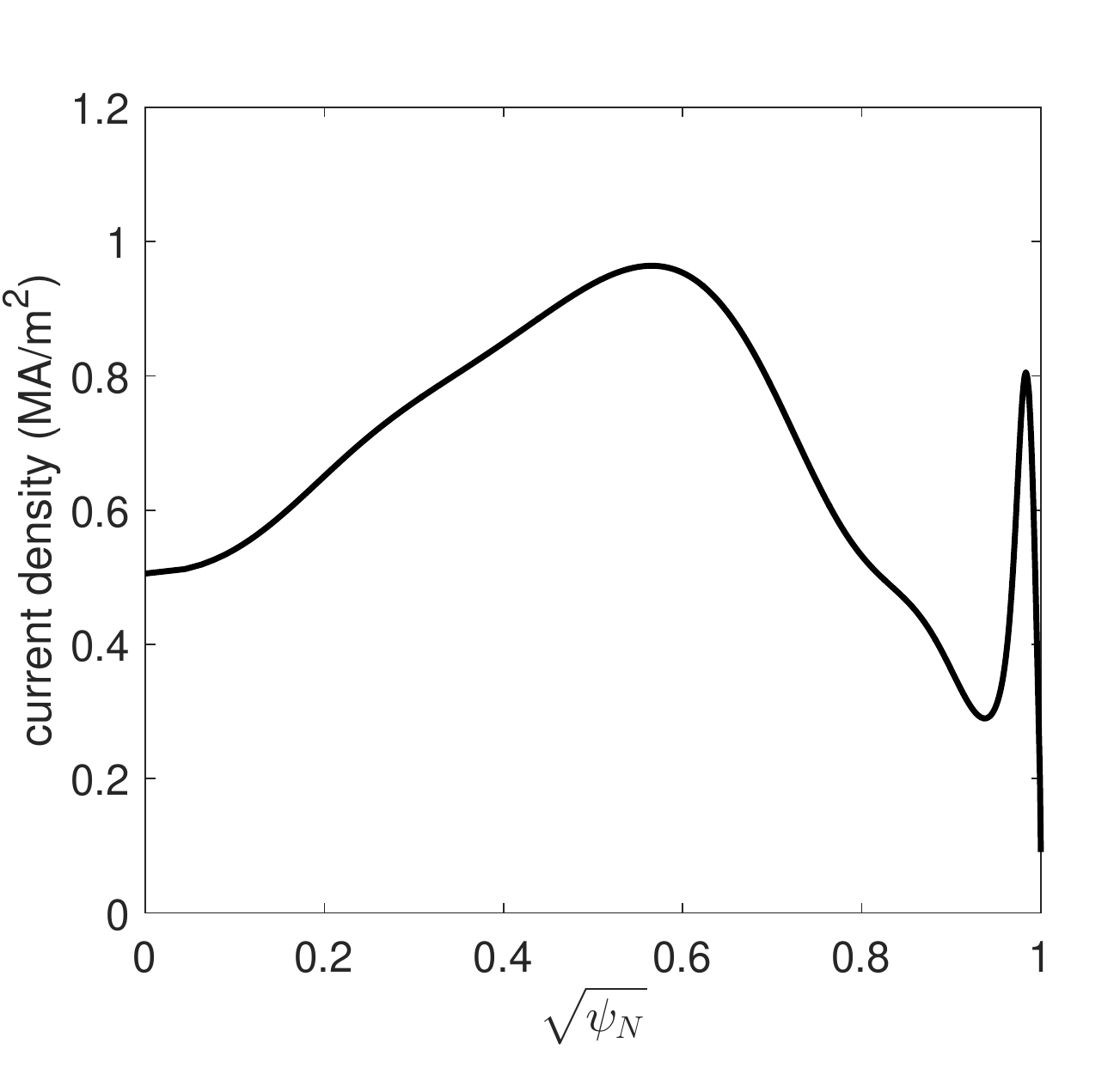}
\put(-220,180){\textbf{(d)}}
\end{minipage}

\caption{Profiles of (a) electron density, (b) ion temperature, (c)safety factor and (d) current density in CFETR baseline equilibrium. $\psi_{\rm N}$ is the normalized poloidal flux function.}
\label{fig_eq} 
\end{figure}

The equilibrium of CFETR baseline scenario considered in our calculation was generated through the integrated modeling in OMFIT framework~\cite{jian2017}. The plasma size is slightly smaller than ITER, with a major radius of $5.7$ m and a minor radius of $1.6$ m. The toroidal magnetic field (5T) and the plasma current (10 MA) at magnetic axis are listed in Table~1 of reference~\cite{jian2017}, among others. Since the baseline case is not designed for demonstrating high fusion gain, the normalized $\beta_N$ is set to $1.88$. Both density (Fig.~\ref{fig_eq}a) and temperature (Fig.~\ref{fig_eq}b) profiles show an edge pedestal region inside the last closed flux surface (LCFS).  Safety factor (q) profile has strong reverse shear region (Fig.~\ref{fig_eq}c) with $q_{\rm min}>2$. The current density profile has highly peaked edge current due to a high fraction of bootstrap current (Fig.~\ref{fig_eq}d). 


\section{MHD model in NIMROD and AEGIS}
\label{sec3}

The MHD equations used in our NIMROD calculations are:
\begin{eqnarray}
\frac{\partial N}{\partial t} + \nabla \cdot \left( N {\bf u}\right) = 0 \\
m N \left( \frac{\partial}{\partial t} + {\bf u} \cdot \nabla \right) {\bf u} = {\bf J} \times {\bf B} - \nabla p - \nabla \cdot {\bf \Pi} \\
\frac{1}{\gamma-1} N \left( \frac{\partial}{\partial t} + {\bf u} \cdot \nabla \right) T = -p\nabla\cdot {\bf u} \\
\frac{\partial {\bf B}}{\partial t} = - \nabla \times \left[ \eta {\bf J} - {\bf u}\times{\bf B}\right] \\
\mu_0 {\bf J} = \nabla \times {\bf B} \qquad \qquad \nabla \cdot {\bf B} = 0
\end{eqnarray}
where {\bf u} is the center-of-mass flow velocity with particle density $N$ and ion mass $m$, ${\bf J}$ is the plasma current, ${\bf B}$ is the magnetic field, $p$ is the plasma pressure, $\eta$ represents the resistivity, ${\bf \Pi}$ is the ion viscous stress tensor, $\gamma$ is the adiabatic index, and $\mu_0$ is the permeability of free space. The initial value NIMROD code has been consistently used in studying different macroscopic phenomena in both fusion and space plasmas~\cite{burke2010,king2016,zhu2013}.

The AEGIS code solves the ideal MHD eigenvalue equation employing the adaptive shooting method along radial direction and the Fourier decomposition in poloidal and toroidal directions. This code has been applied before in evaluating the stability of ideal MHD modes in presence of either conducting or resistive wall for ITER~\cite{zheng17qa,zheng2005}. In AEGIS, the ideal MHD model is used for the plasma region within separatrix, and the vacuum region extends from separatrix to the first wall. On the contrary, NIMROD uses the resistive MHD model for both the plasma within separatrix and the low temperature plasma of vacuum-like halo region between separatrix and the first wall. 


\section{Dominant low-$n$ ideal MHD instabilities in presence of perfect conducting wall}
\label{sec4}

For the purpose of benchmark and comparison, a step-like hyperbolic tangent resistivity profile is adopted in NIMROD to model the ideal core plasma and the vacuum-like halo region, where the Lundquist number (defined as $S=\tau_{\rm R}/\tau_{\rm A}$) of core plasma region $S_{plasma} = 1.146\times10^{10}$ and halo region $S_{halo} = 1.146\times10^{8}$. Here, $\tau_{\rm R} = \mu_0 a^2/\eta$ is the resistive diffusion time with $a$ being the minor radius and $\tau_{\rm A} = R_0 \sqrt{\mu_0\rho_{m0}}/B_0$ is the Alfv\'en time with $R_0$ being the major radius of the magnetic axis, $B_0$ and $\rho_{m0}$ the values of magnetic field and mass density at magnetic axis respectively. $S_{plasma}$ and $S_{halo}$ have been scanned to identify their asymptotic values for the ideal MHD regime in plasma and the vacuum regime in halo regions, respectively, which are able to yield the converged value of growth rate in the ideal MHD limit.

\begin{figure}[htbp]
\centering
\begin{minipage}{0.49\textwidth}
\includegraphics[width=1.0\textwidth]{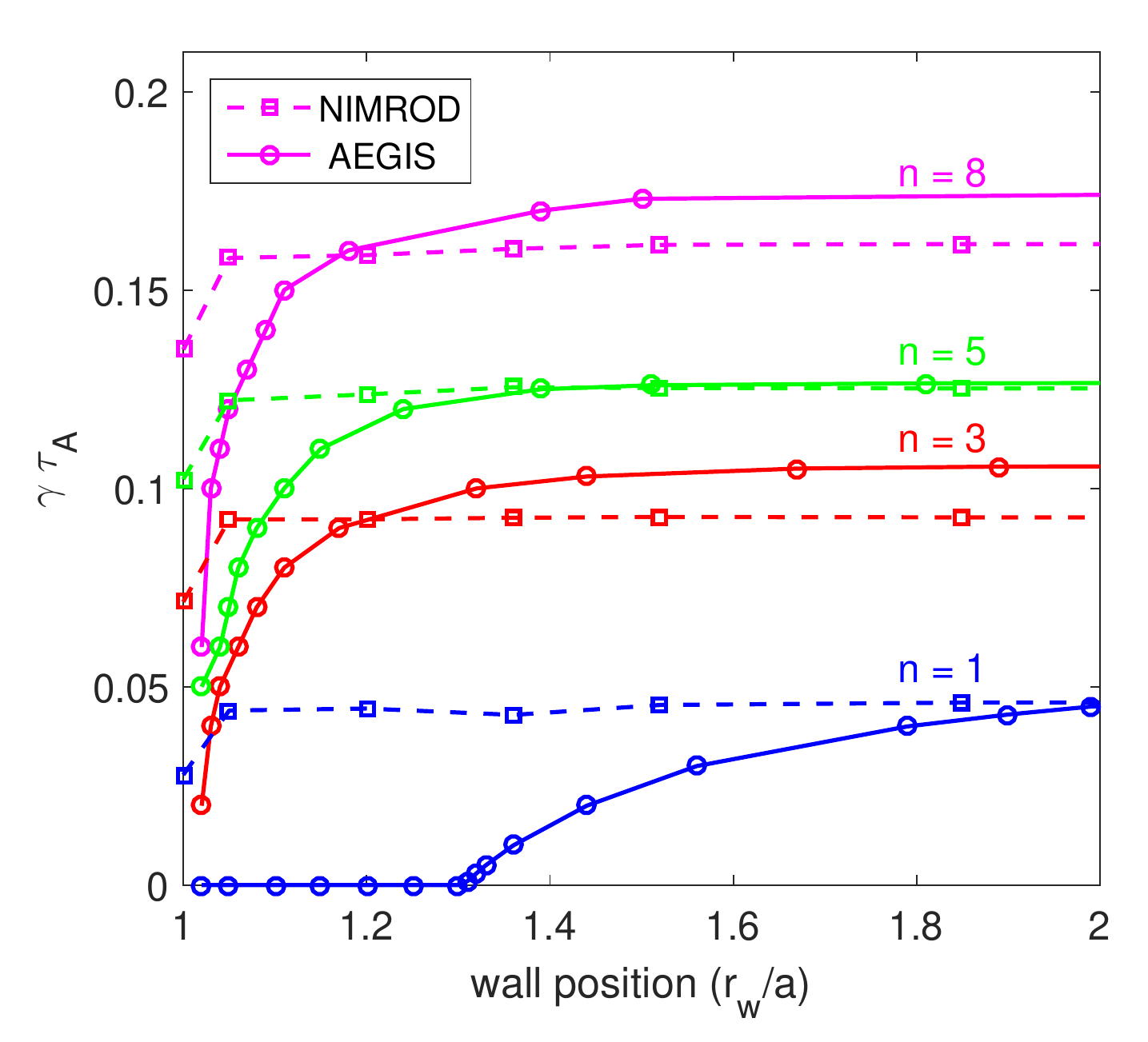}
\put(-220,180){\textbf{(a)}}
\end{minipage}
\begin{minipage}{0.49\textwidth}
\includegraphics[width=1.0\textwidth]{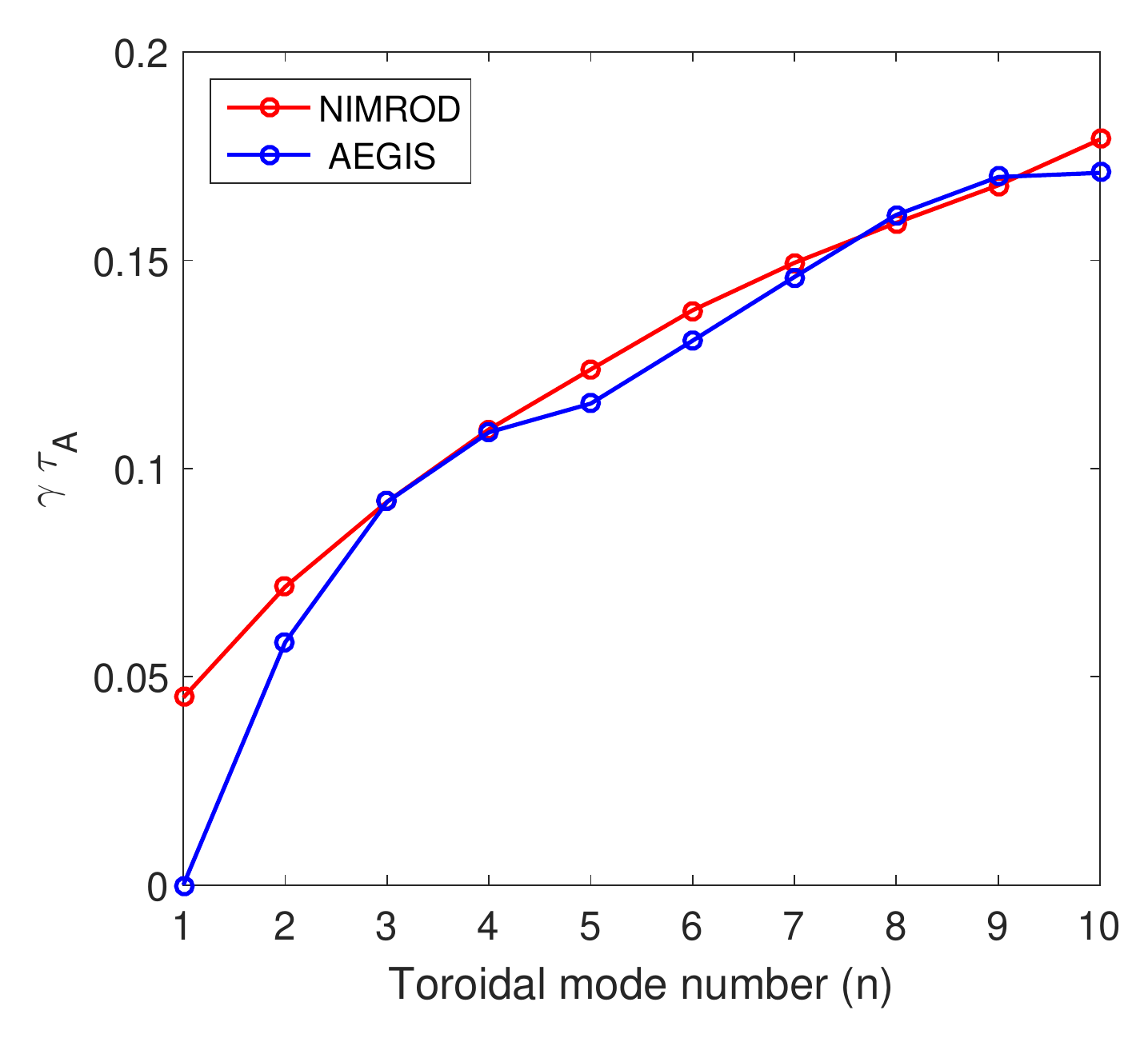}
\put(-220,180){\textbf{(b)}}
\end{minipage}

\caption{(a) Ideal MHD growth rates as functions of the perfect conducting wall location for $n= 1,3,5,8$, from NIMROD and AEGIS calculations. (b) Ideal MHD growth rates versus toroidal number $n$ with perfect conducting wall at position $r_w=1.2a$ from NIMROD and AEGIS calculations. }
\label{fig_benchmark} 
\end{figure}

Employing this resistivity model, we are able to study the $n=1-10$ ideal MHD modes using NIMROD. In Fig.~\ref{fig_benchmark}a, the $n = 1,3,5,8$ ideal MHD growth rates are calculated for a range of perfect conducting wall locations ($r_w$), from being close to LCFS to $r_w=2a$. We find that the growth rates of all $n$ modes reach the no-wall limit at a close proximity to LCFS, suggesting a weak dependence on the conducting wall position. This finding differs from AEGIS, where ideal wall limit is reached only far away from plasma boundary, especially for lower-$n$ modes (Fig.~\ref{fig_benchmark}a). For the designed wall location $r_w = 1.2a$, AEGIS (NIMROD) finds $n = 1$ mode stable (unstable). However, good agreement is achieved for growth rates of all $n = 2-9$ modes between AEGIS and NIMROD. We note that in NIMROD calculation the X-point is included in computation domain. However in AEGIS calculation, the plasma region is truncated at the 99\% poloidal flux surface ($q = 4.15$), which may possibly be the reason for the difference in growth rates. 

\begin{figure}[htbp]
\centering
\begin{minipage}{0.49\textwidth}
\includegraphics[width=1.0\textwidth]{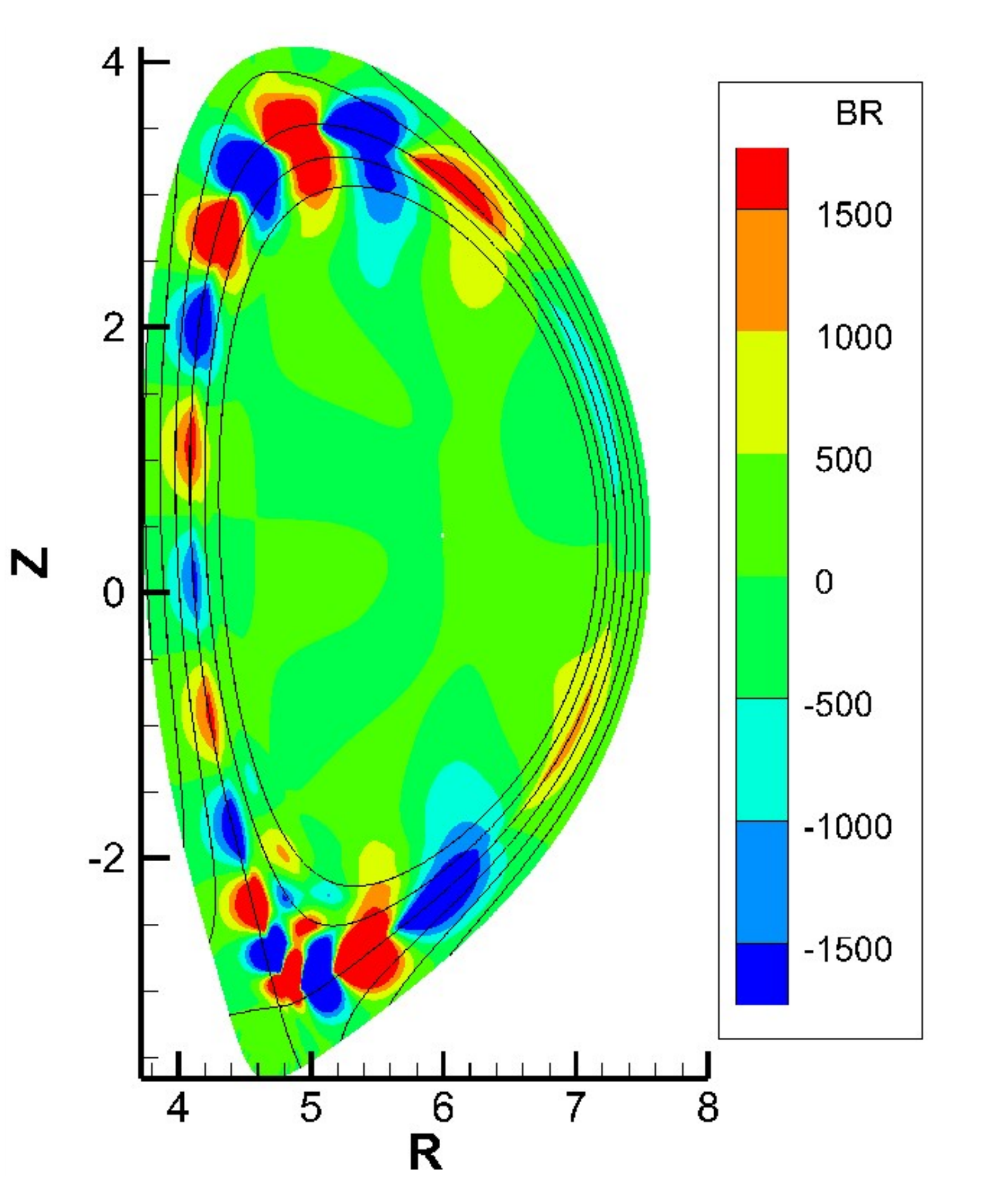}
\put(-220,240){\textbf{(a)}}
\end{minipage}
\begin{minipage}{0.49\textwidth}
\includegraphics[width=1.0\textwidth]{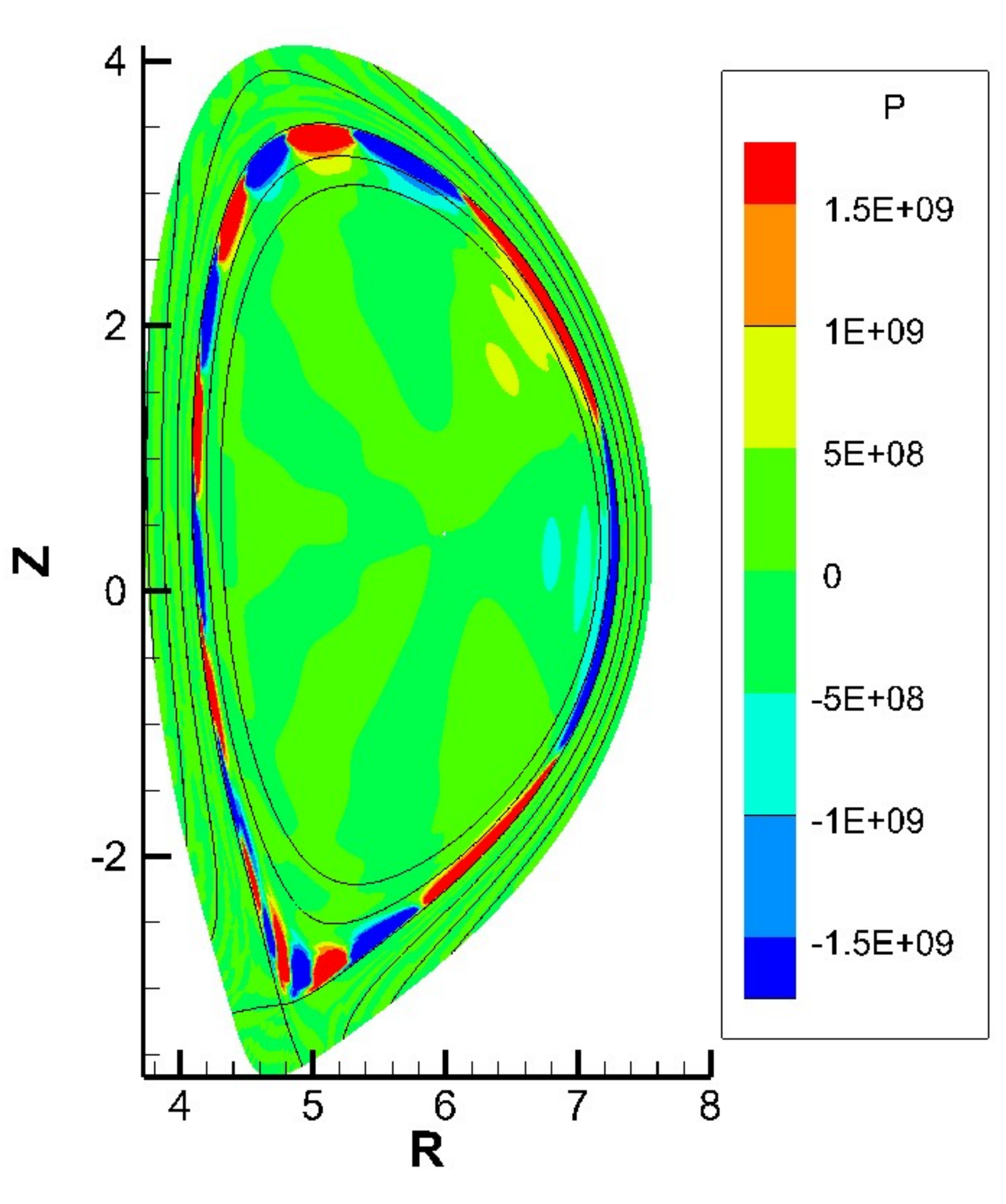}
\put(-220,240){\textbf{(b)}}
\end{minipage}

\begin{minipage}{0.49\textwidth}
\includegraphics[width=1.0\textwidth]{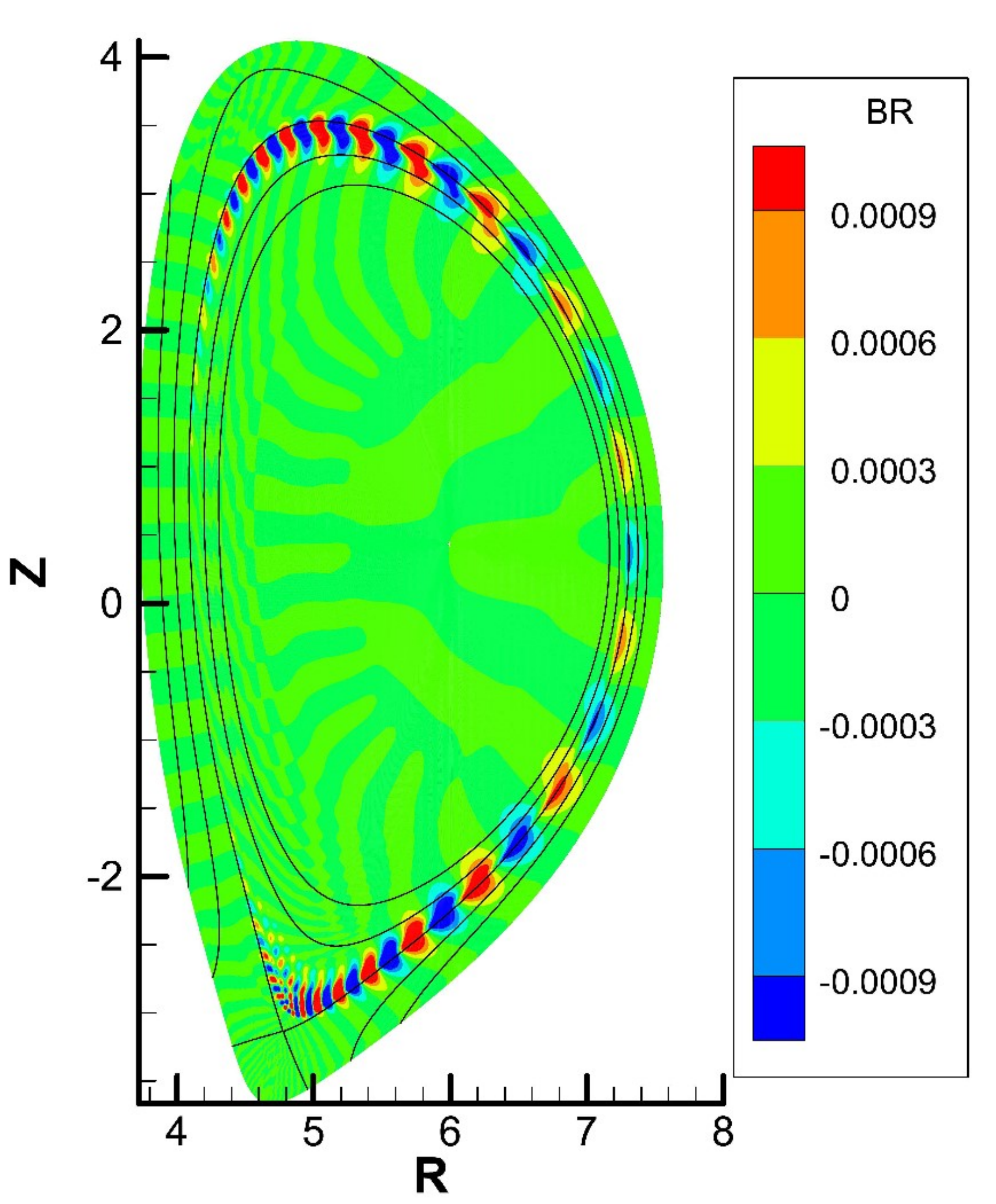}
\put(-220,240){\textbf{(c)}}
\end{minipage}
\begin{minipage}{0.49\textwidth}
\includegraphics[width=1.0\textwidth]{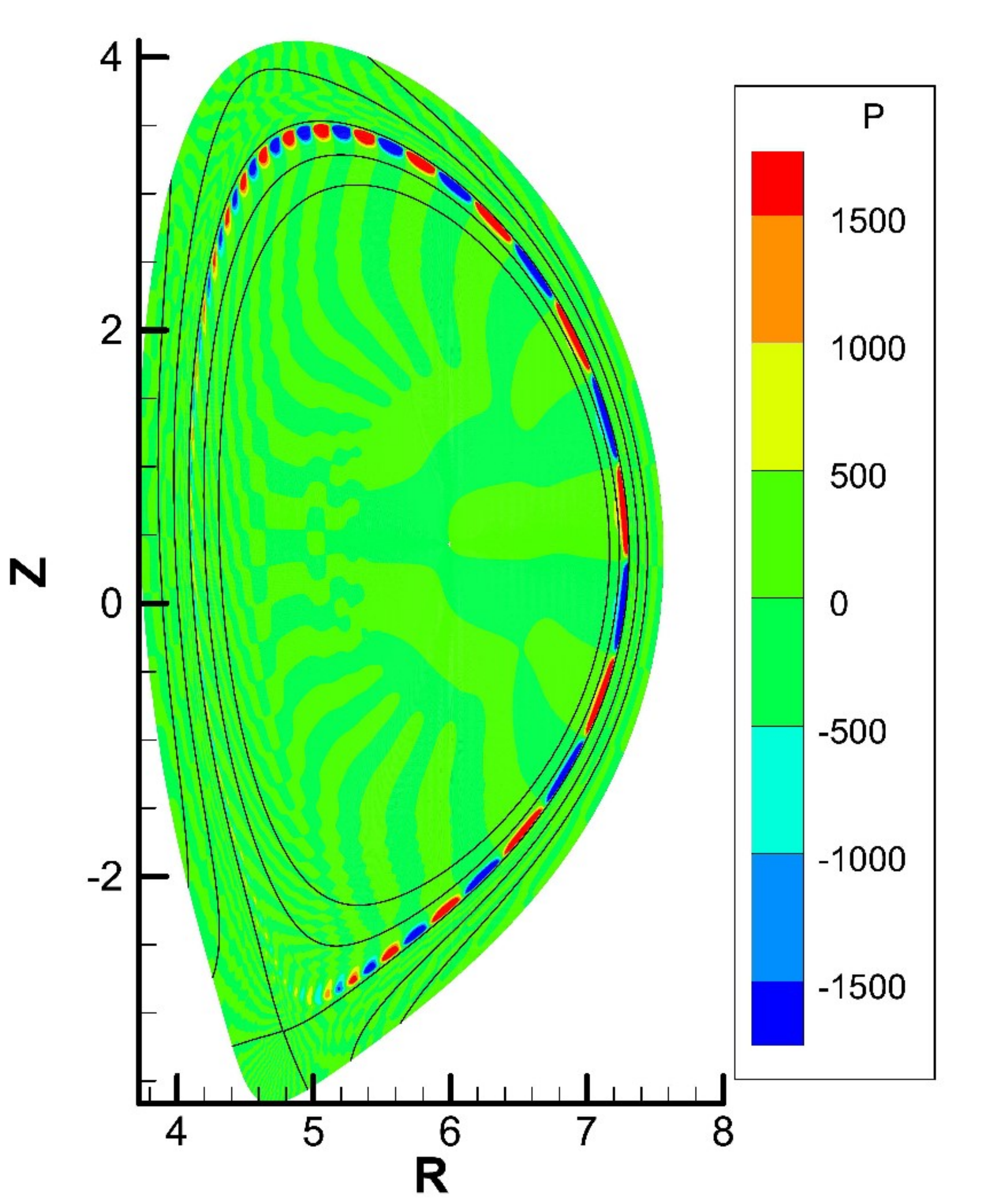}
\put(-220,240){\textbf{(d)}}
\end{minipage}

\caption{Dominant linear mode structures in presence of hyperbolic tangent resistivity profile and self-similar conducting wall at position $r_w=1.2a$ as shown in the color contours of: (a) radial component of perturbed magnetic field ($B_r$), and (b) perturbed pressure (P) of $n=2$ mode, and (c) perturbed $B_r$ of $n=8$ mode, and (d) perturbed pressure of $n=8$ mode. The poloidal magnetic flux of the equilibrium is shown as the dark line contours in each plot.}
\label{fig_contour_ideal}  
\end{figure}

\begin{figure}[htbp]
\centering
\begin{minipage}{0.49\textwidth}
\includegraphics[width=1.0\textwidth]{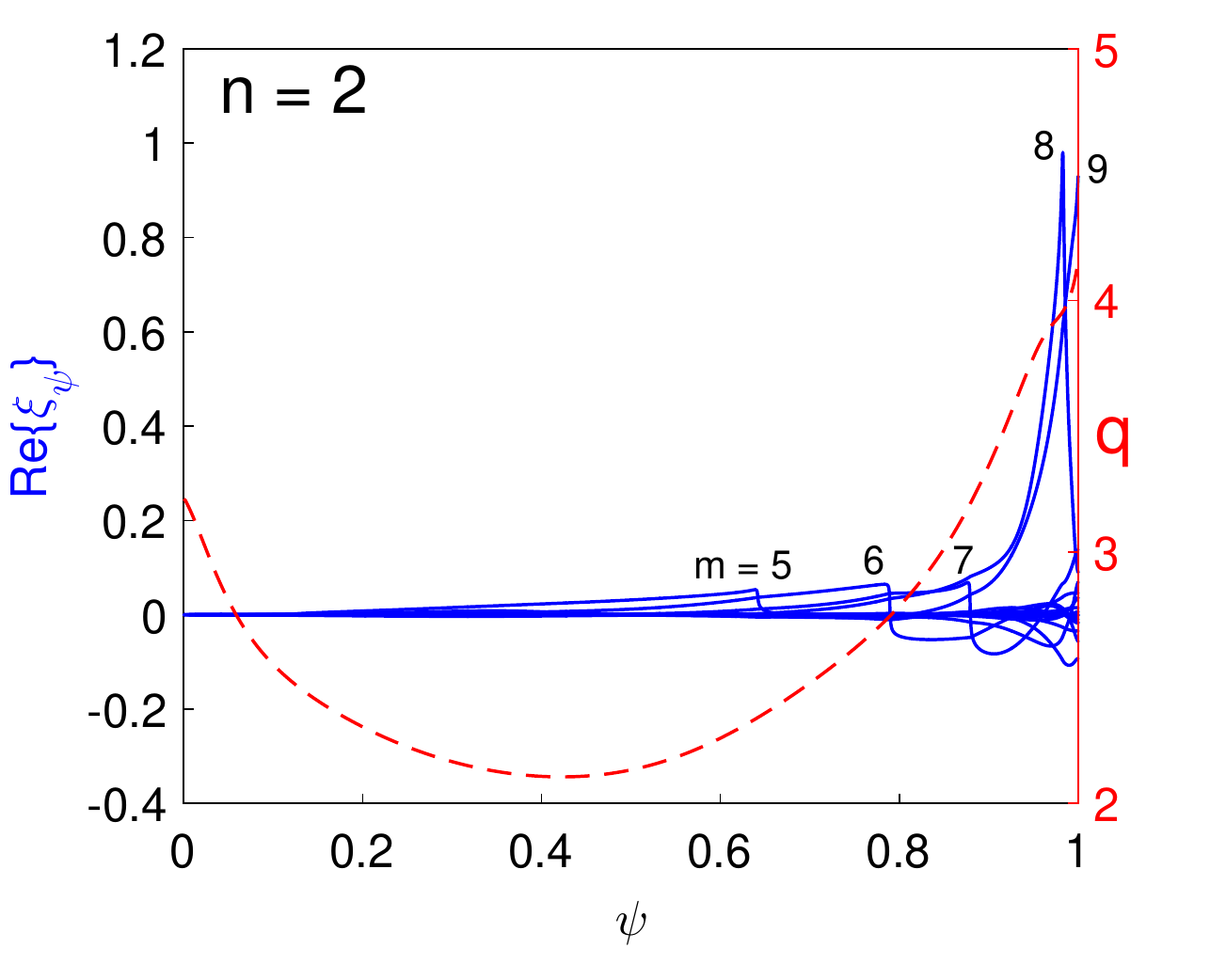}
\put(-220,150){\textbf{(a)}}
\end{minipage}
\begin{minipage}{0.49\textwidth}
\includegraphics[width=1.0\textwidth]{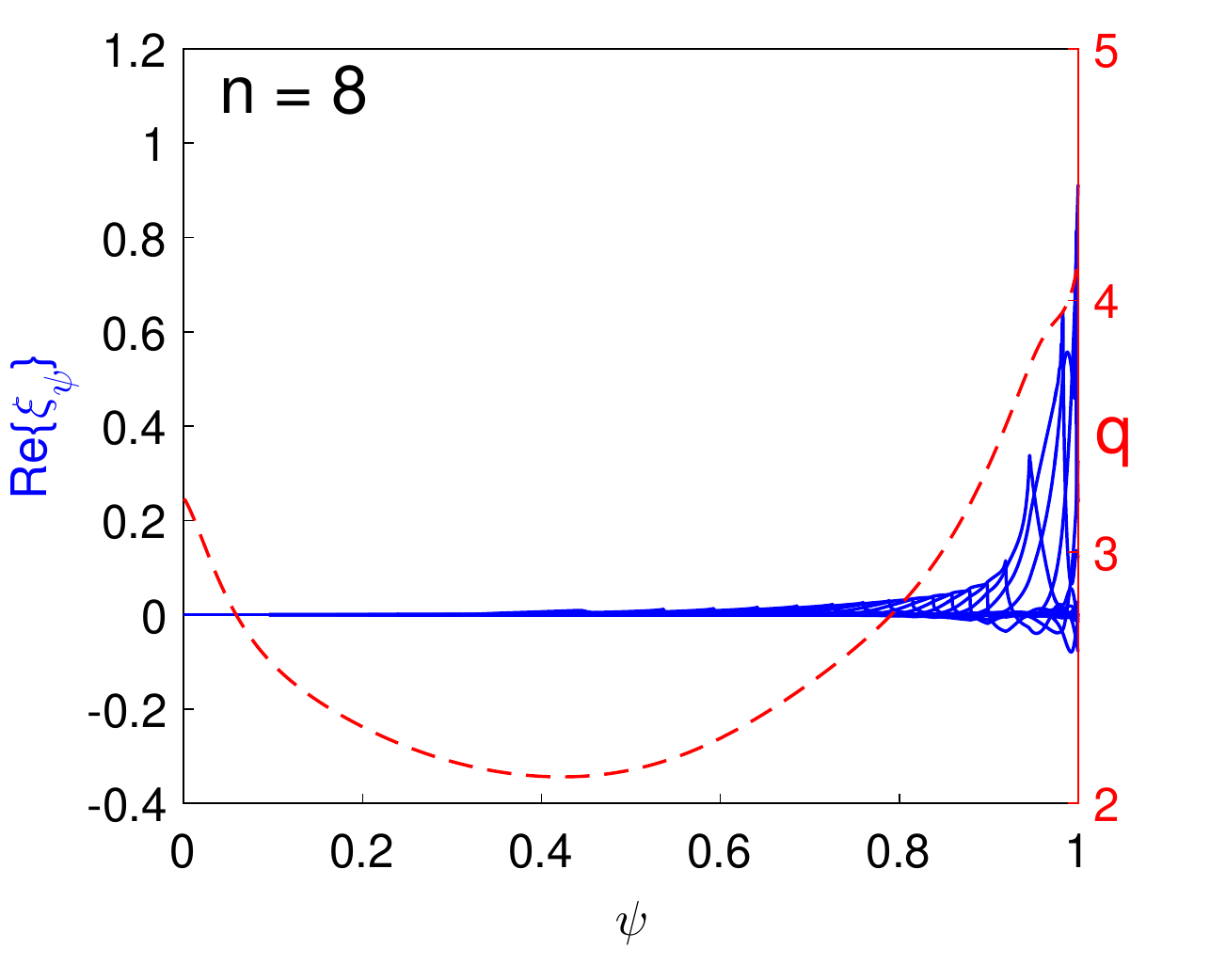}
\put(-220,150){\textbf{(b)}}
\end{minipage}

\caption{Real component of radial displacements for (a) $n = 2$ and (b) $n = 8$ ideal MHD modes with perfect conducting wall at position $r_w=1.2a$, respectively.}
\label{fig_eigen_ideal} 
\end{figure}

The variations in mode structure for different toroidal number $n$ are also consistent between NIMROD (Fig.~\ref{fig_contour_ideal}) and AEGIS (Fig.~\ref{fig_eigen_ideal}). For the perturbed magnetic field and pressure from NIMROD calculations, the mode structures in poloidal plane are apparently different between the $n=2$ mode and the $n=8$ mode (Fig.~\ref{fig_contour_ideal}). The $n=2$ mode structure is broader and the magnetic field perturbation extends well into the vacuum region across separatrix. whereas the $n=8$ mode structure is much narrower and localized in the pedestal region inside separatrix. For the perturbed radial displacements from AEGIS calculations, the two dominant components $m = 8$ and $9$ of the $n=2$ mode are peaked near the edge region, and the $m = 9$ component has the typical profile of an external kink mode (Fig.~\ref{fig_eigen_ideal}a). For $n = 8$ mode (Fig.\ref{fig_eigen_ideal}b), all $m$ components are localized in the pedestal region near edge.


\section{Effects of realistic plasma-vacuum profile and wall shape}
\label{sec5}

\begin{figure}[htbp]
\centering
\begin{minipage}{0.49\textwidth}
\includegraphics[width=1.0\textwidth]{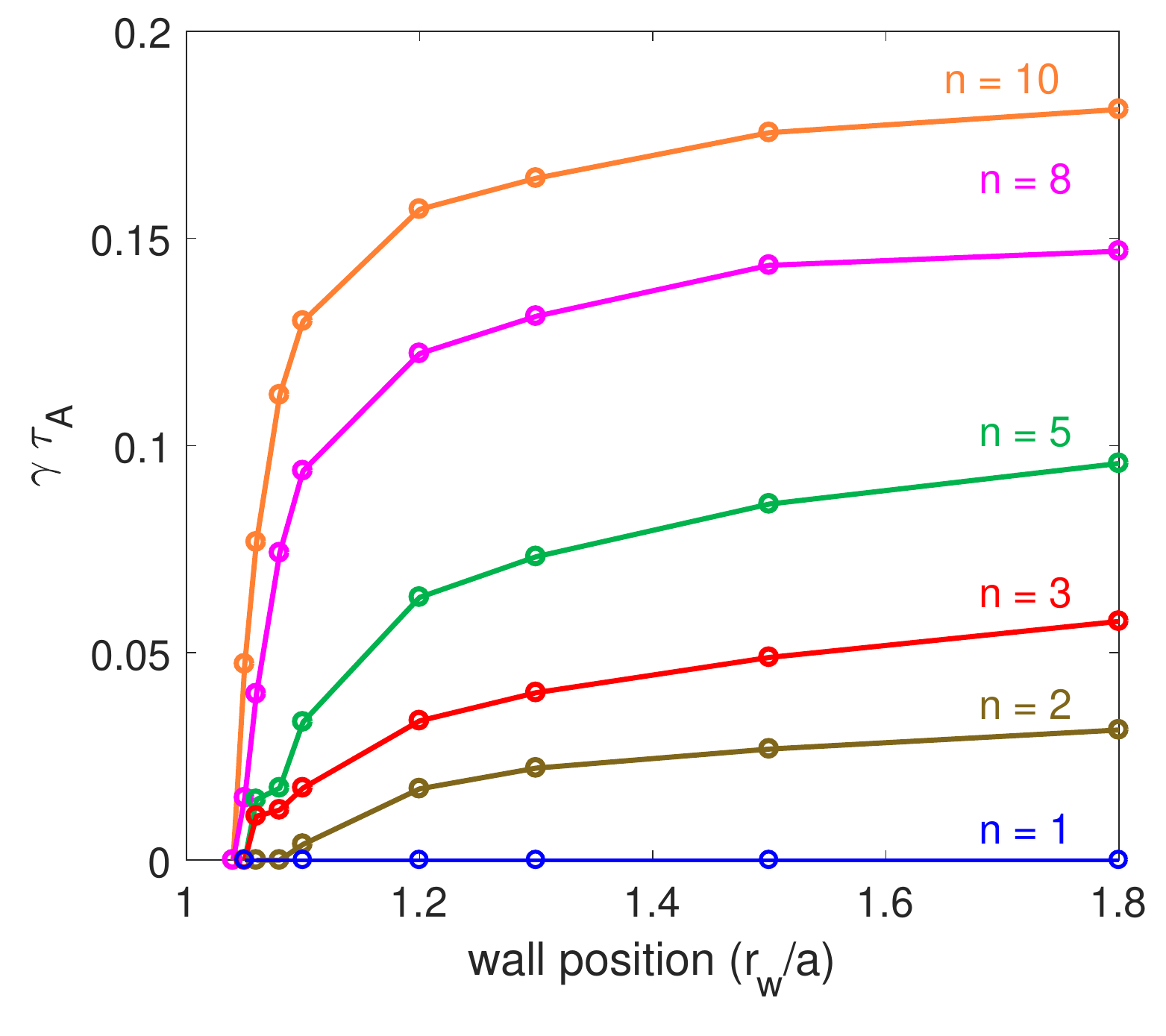}
\put(-220,180){\textbf{(a)}}
\end{minipage}
\begin{minipage}{0.49\textwidth}
\includegraphics[width=1.0\textwidth]{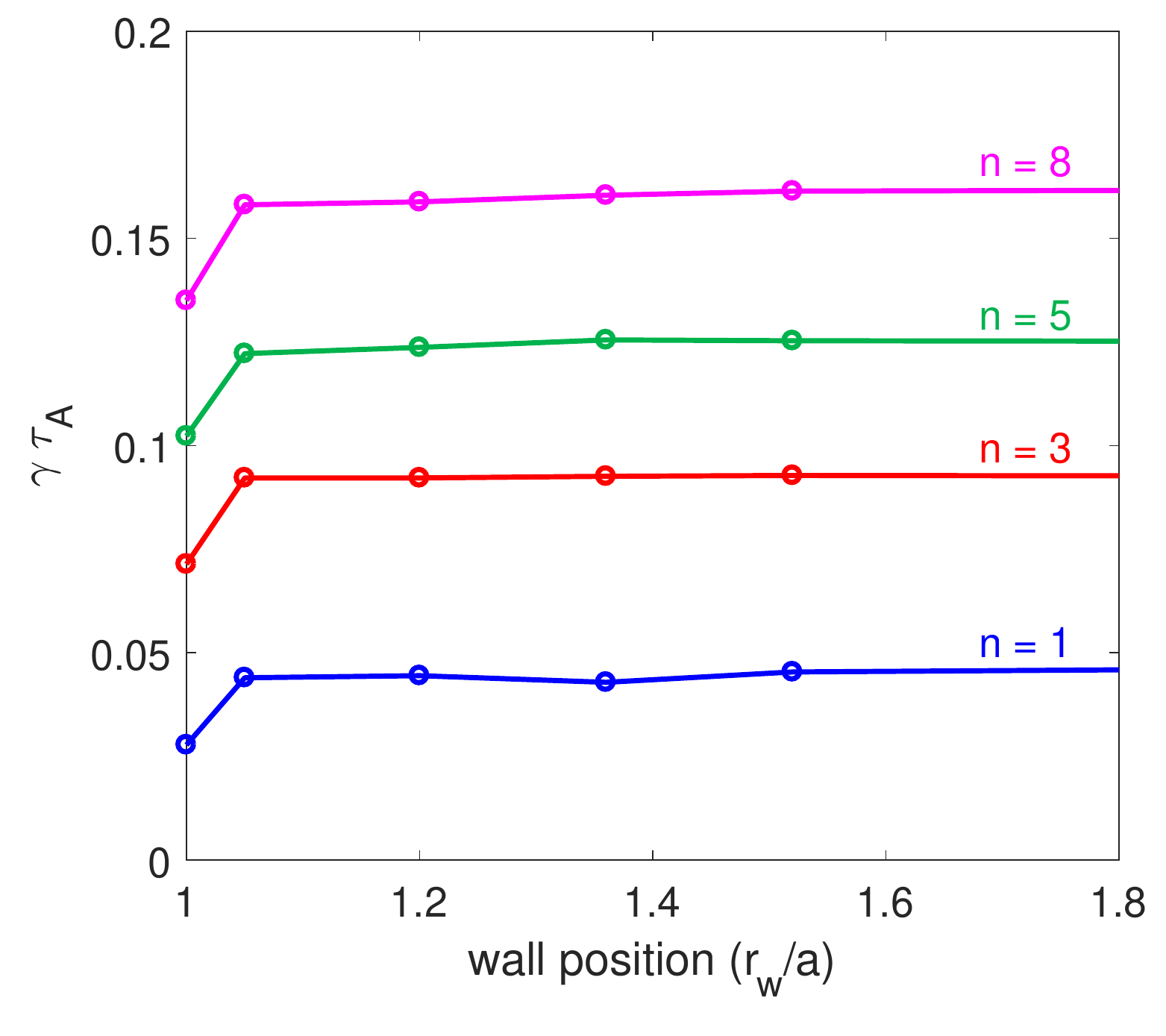}
\put(-220,180){\textbf{(b)}}
\end{minipage}

\caption{Low-$n$ global mode growth rates as functions of the perfect conducting wall location for different toroidal numbers using plasma-vacuum profiles of (a) Spitzer and (b) step-like hyperbolic tangent function resistivity models, respectively, from NIMROD calculations.}
\label{fig_spitzer}
\end{figure}

The stability of $n = 1 - 10$ modes has been re-evaluated using NIMROD after considering the more realistic resistivity profile based on the Spitzel model, i.e. $\eta = \eta_0 (T_{{\rm e}0}/T_{\rm e})^{3/2}$, where $T_{{\rm e}0}$, $\eta_0$, $T_{\rm e}$ denote the electron temperature, the resistivity at magnetic axis, and the electron temperature profile respectively. The normalized ideal MHD growth rates of $n=1,2,3,5,8,10$ are plotted in the Fig.~\ref{fig_spitzer} with a self-similar wall position varying from close to separatrix to $r_w=1.8a$. A particular wall position $r_w=1.04a$ is identified where no mode is found unstable inside. The growth rates of all modes increase rapidly until the wall position $1.2a$ is reached. Afterwards, they gradually approach their corresponding no-wall limit values. The wall positions for all modes transitioning to no-wall limit are basically the same, and the growth rate at no-wall limit increases monotonically with mode numbers from $n=2$ to $n=10$. Presence of Spitzer resistivity profile stabilizes the $n=1$ mode which is found unstable using the hyperbolic tangent resistivity profile (Fig.~\ref{fig_spitzer} and \ref{fig_benchmark}a).

\begin{figure}[htbp]
\centering
\begin{minipage}{0.49\textwidth}
\includegraphics[width=1.0\textwidth]{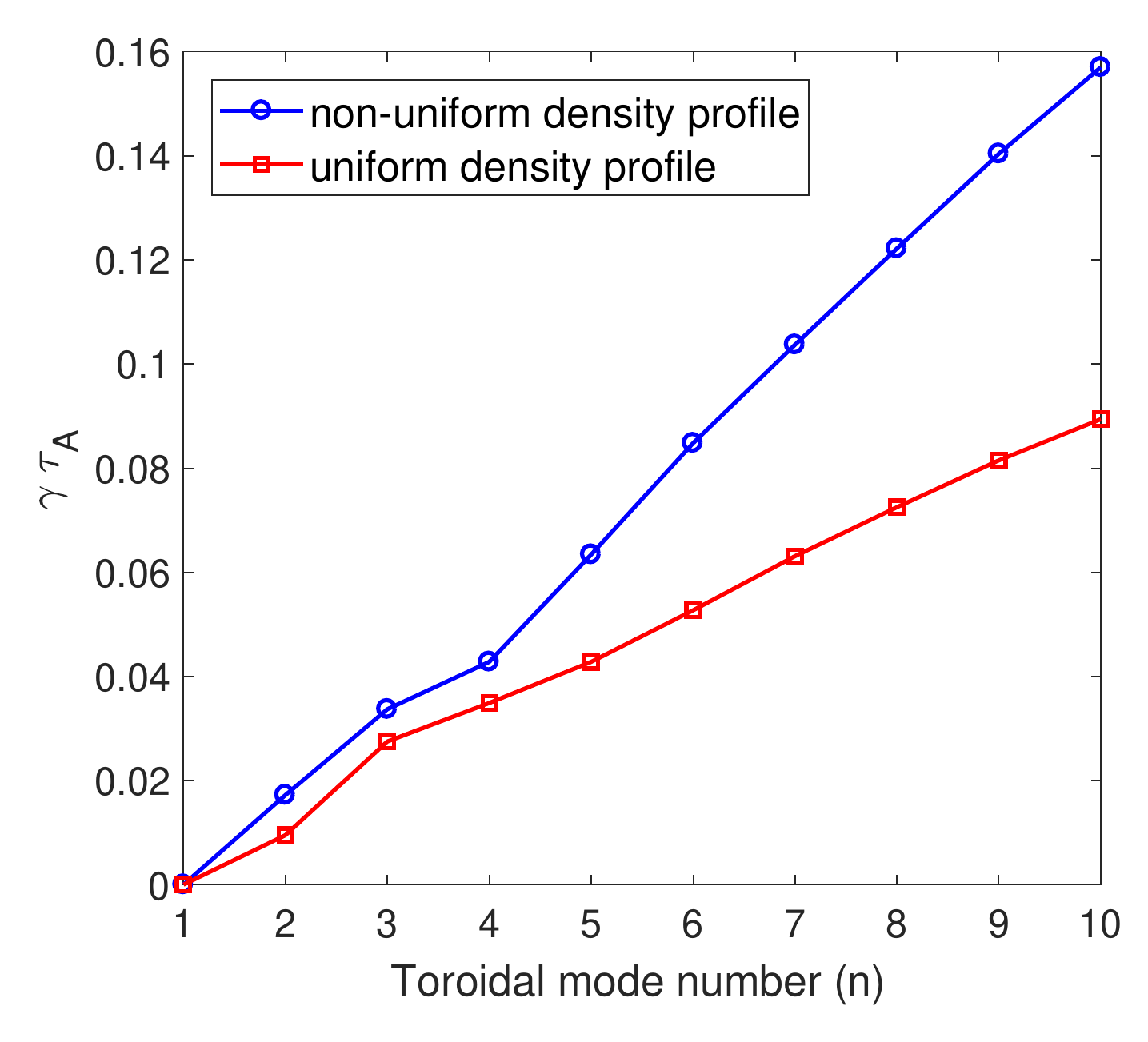}
\put(-220,180){\textbf{(a)}}
\end{minipage}
\begin{minipage}{0.49\textwidth}
\includegraphics[width=1.0\textwidth]{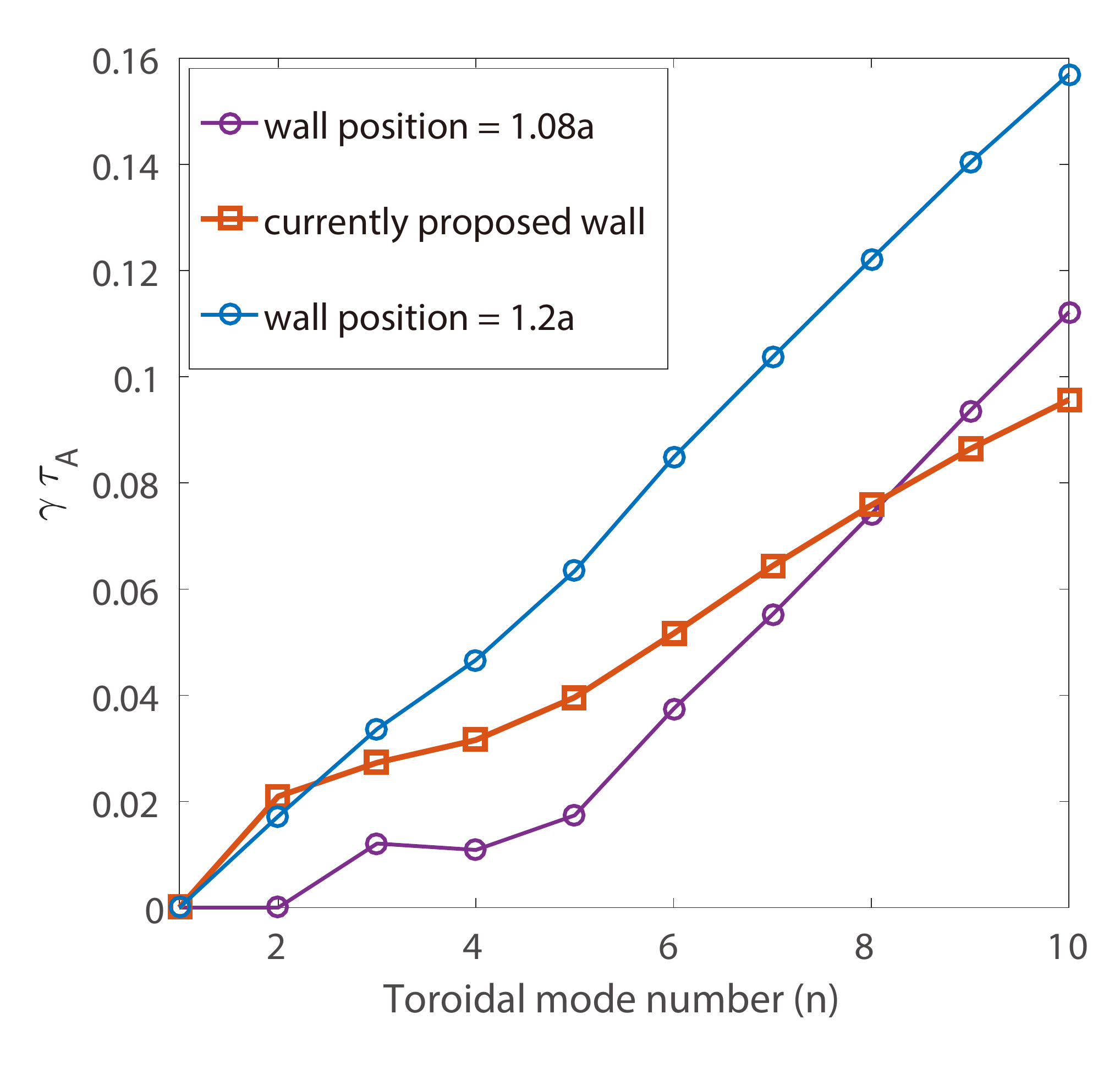}
\put(-220,180){\textbf{(b)}}
\end{minipage}

\caption{Growth rates of ideal MHD modes as functions of toroidal mode number $n$ (a) with perfect conducting wall at position $r_w = 1.2a$ for uniform and non-uniform density profiles, and (b) for different shapes of perfect conducting wall.}
\label{fig_nd_realwall}
\end{figure}

All above numerical results from both NIMROD and AEGIS calculations are based on non-uniform density profiles with high gradient in pedestal region. The density pedestal gradient has driven the $n=2-10$ modes more unstable than the uniform density case (Fig.~\ref{fig_nd_realwall}a). The higher the toroidal mode number is, the stronger is the influence of density pedestal on growth rate. Here, level of uniform density is kept same as the value of density profile at magnetic axis, therefore the normalizing Alfv\'en time scale ($\tau_ {\rm A}=6.627 \times 10^{-7} {\rm s}$) is same for both density cases.

\begin{figure}[htbp]
\centering
\begin{minipage}{0.49\textwidth}
\includegraphics[width=1.0\textwidth]{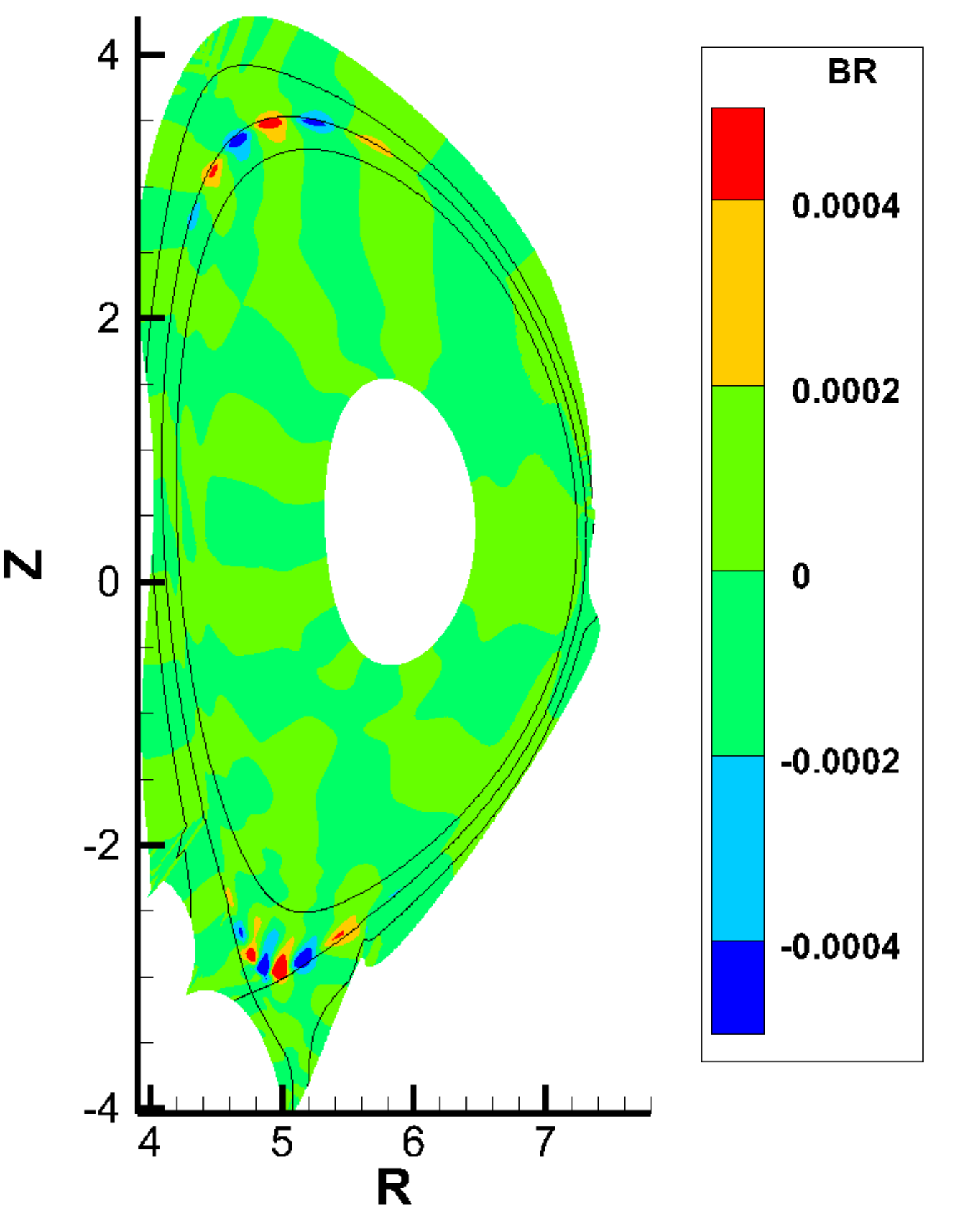}
\put(-220,240){\textbf{(a)}}
\end{minipage}
\begin{minipage}{0.49\textwidth}
\includegraphics[width=1.0\textwidth]{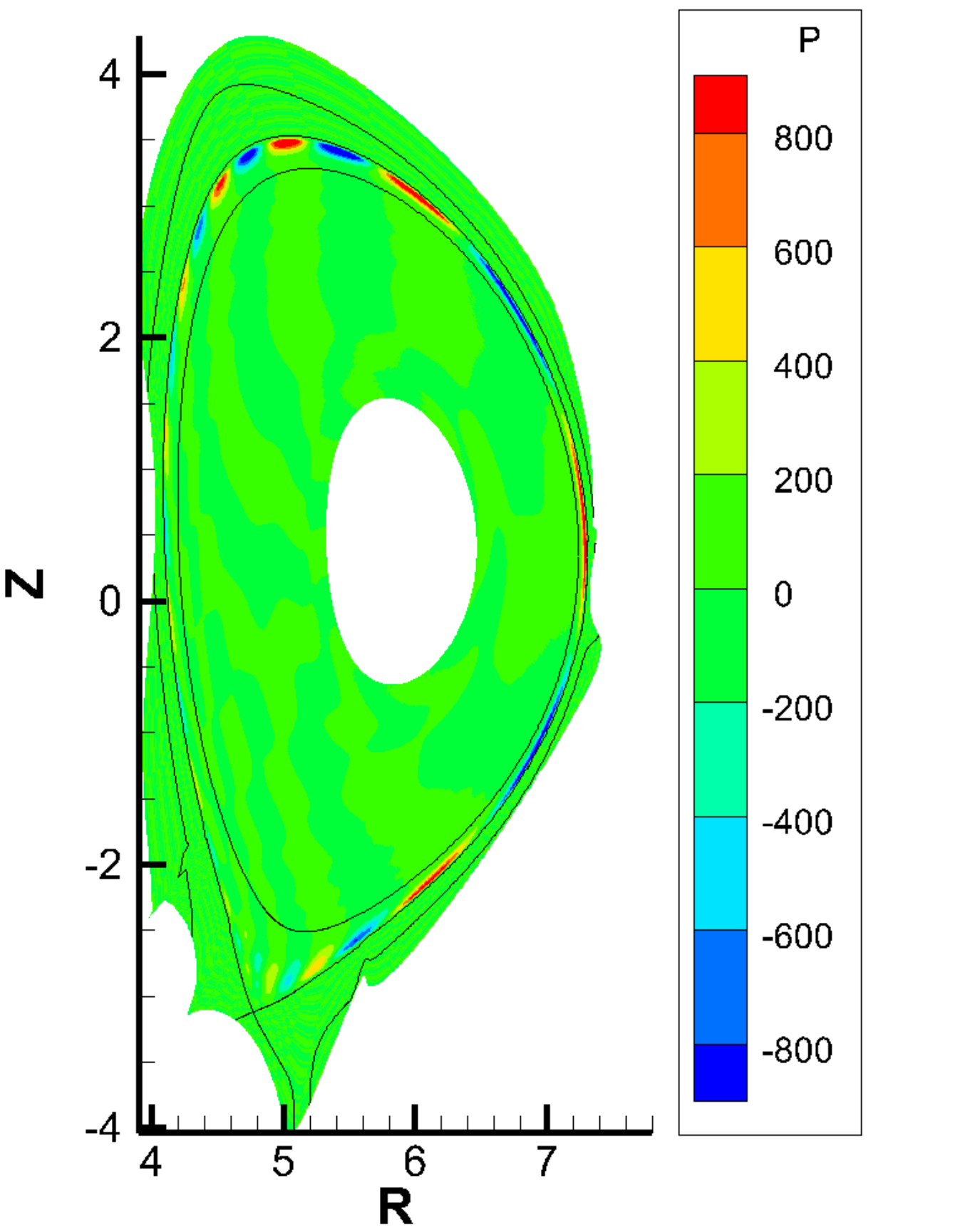}
\put(-220,240){\textbf{(b)}}
\end{minipage}

\begin{minipage}{0.49\textwidth}
\includegraphics[width=1.0\textwidth]{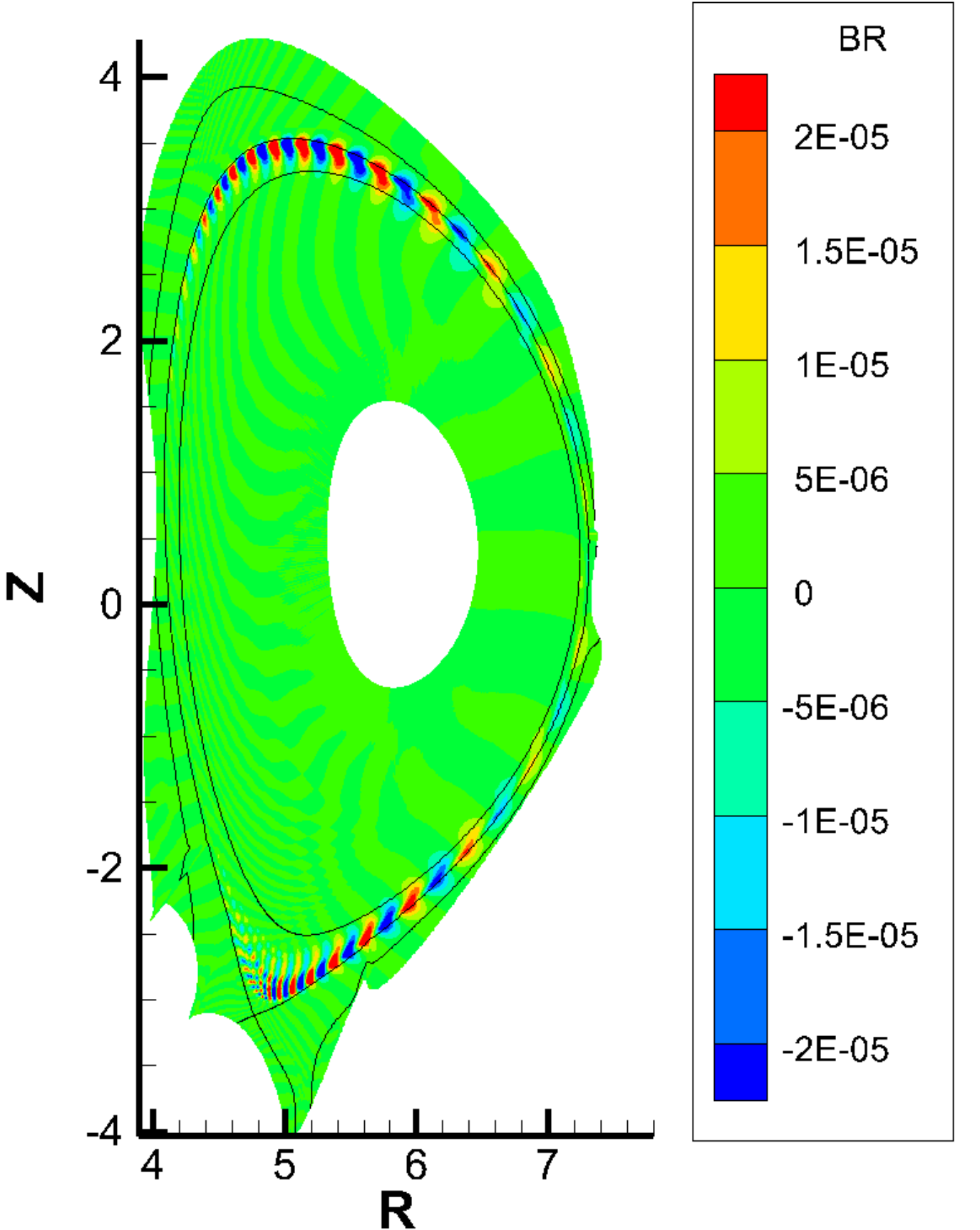}
\put(-220,240){\textbf{(c)}}
\end{minipage}
\begin{minipage}{0.49\textwidth}
\includegraphics[width=1.0\textwidth]{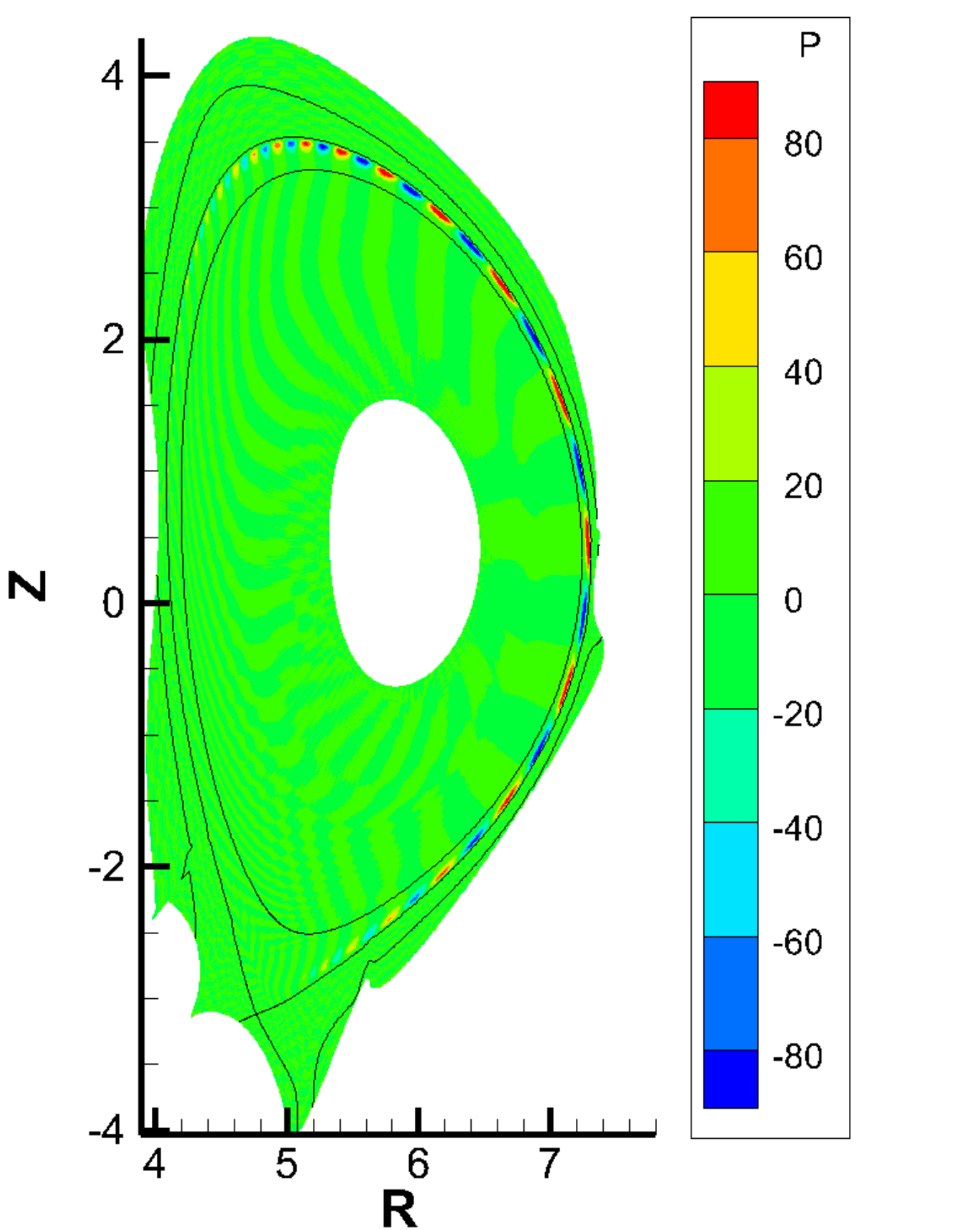}
\put(-220,240){\textbf{(d)}}
\end{minipage}

\caption{Dominant linear mode structures in presence of Spitzer resistivity profile and the proposed real first-wall as shown in the color contours of: (a) perturbed $B_r$ of $n=3$ mode, and (b) perturbed pressure of $n=3$ mode, and (c) perturbed $B_r$ of $n=10$ mode, and (d) perturbed pressure of $n=10$ mode. The proposed real first-wall configuration is shown as the boundary of the contours.}
\label{fig_con_realwall}
\end{figure}

The growth rate calculations of different modes have also been carried out after considering recently proposed real first-wall configuration of CFETR (shown in Fig.~\ref{fig_con_realwall} as the boundary of contour). The designed first-wall position is near the wall location of $r_w=1.2a$, but shape is different from self-similar wall often used in MHD stability calculations. A stabilizing effect of real first-wall shape is noticed for higher-$n$ modes, whose growth rates are close to those with self-similar wall at $r_w=1.08a$, whereas for low-$n$ modes, their growth rates are similar to the self-similar wall at $r_w=1.2a$ (Fig.~\ref{fig_nd_realwall}b).

\begin{figure}[htbp]
\centering
\begin{minipage}{0.49\textwidth}
\includegraphics[width=1.0\textwidth]{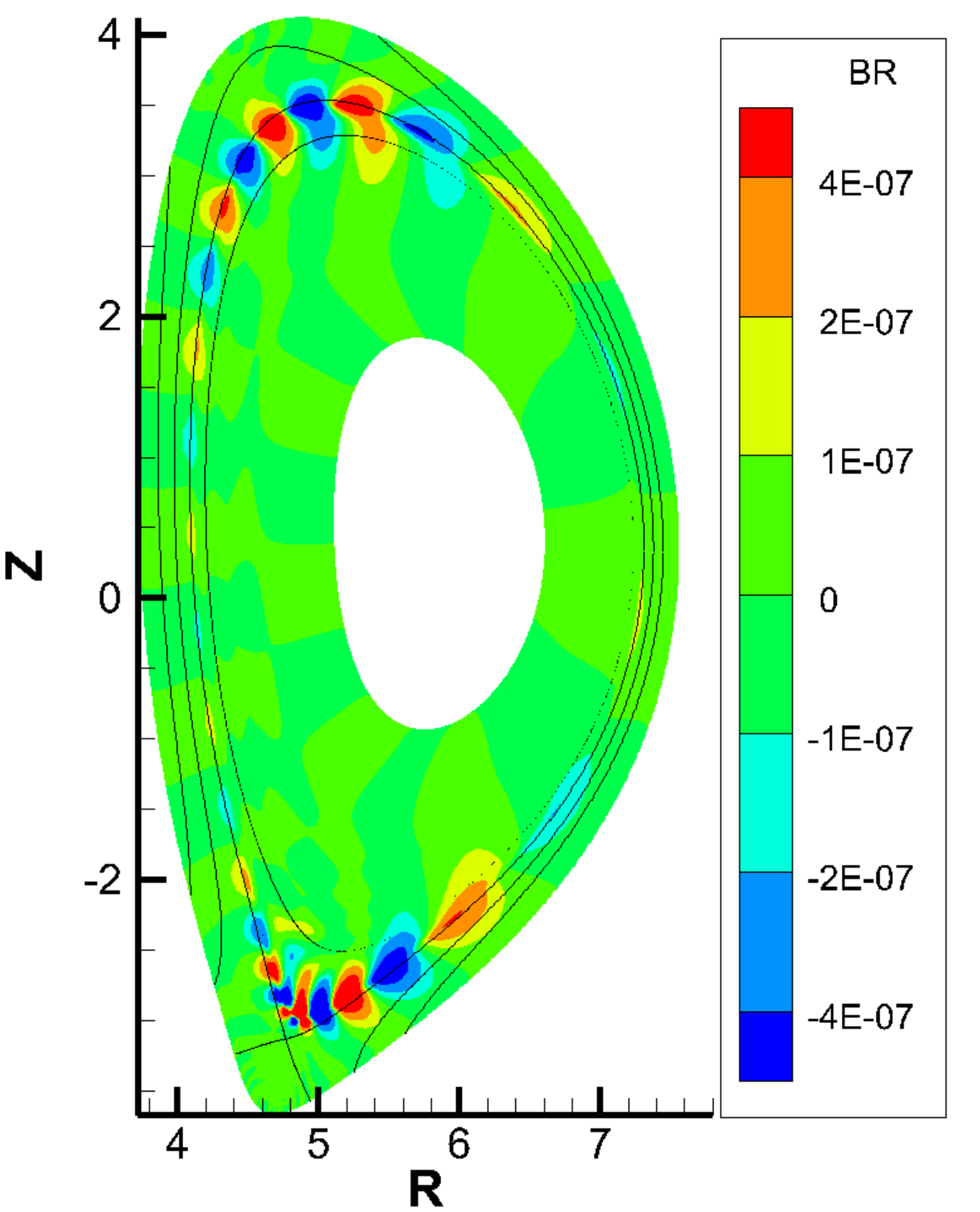}
\put(-220,240){\textbf{(a)}}
\end{minipage}
\begin{minipage}{0.49\textwidth}
\includegraphics[width=1.0\textwidth]{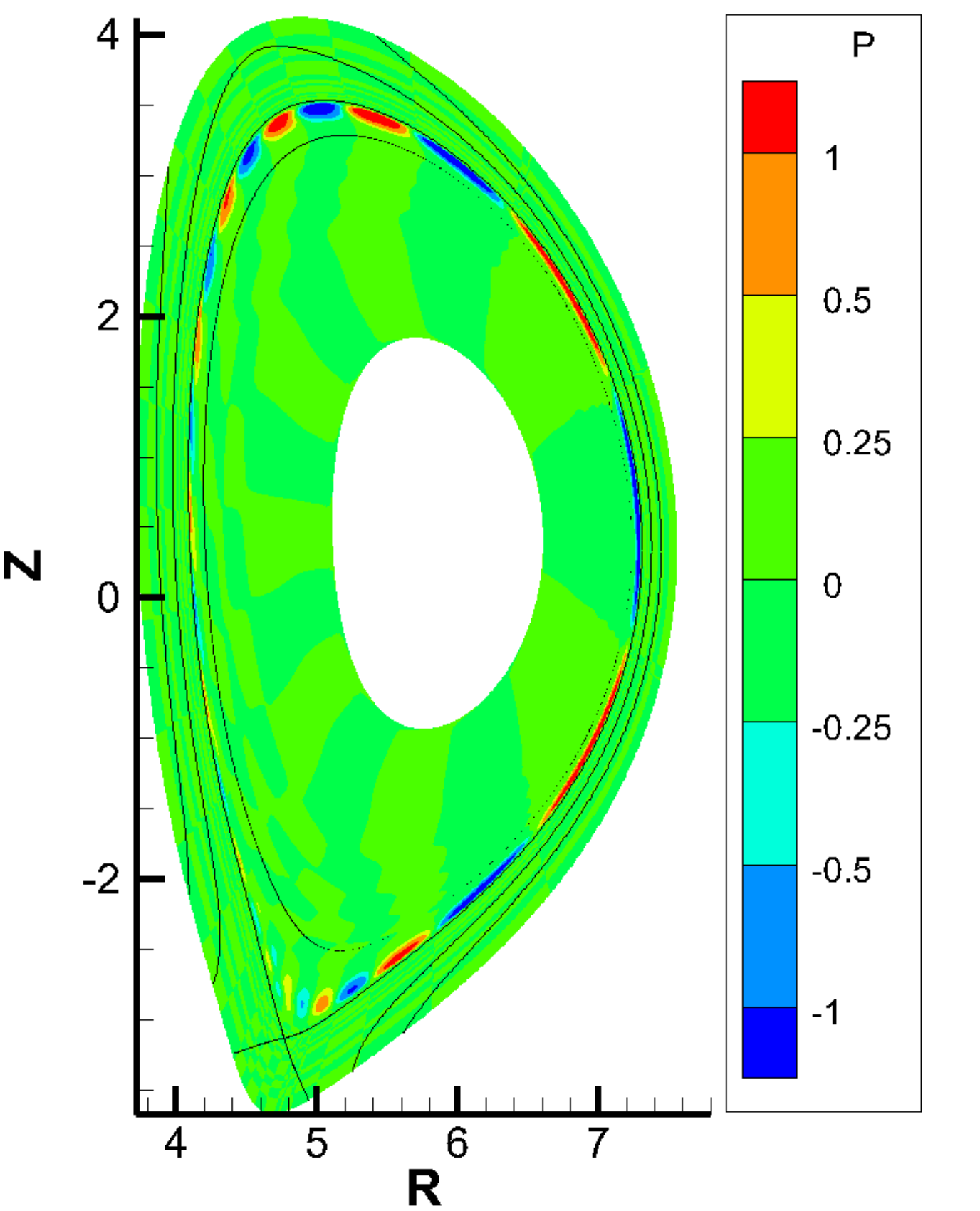}
\put(-220,240){\textbf{(b)}}
\end{minipage}

\begin{minipage}{0.49\textwidth}
\includegraphics[width=1.0\textwidth]{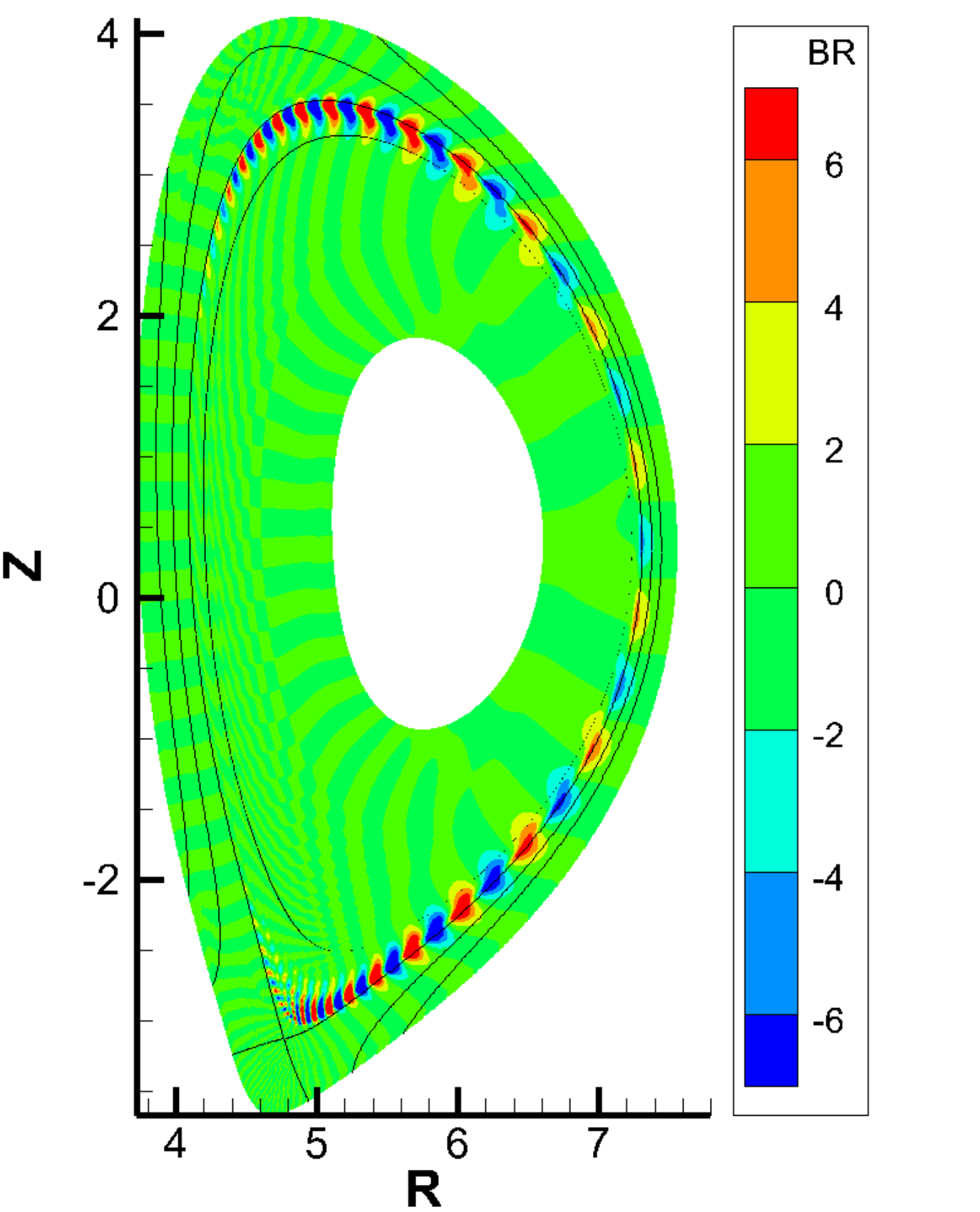}
\put(-220,240){\textbf{(c)}}
\end{minipage}
\begin{minipage}{0.49\textwidth}
\includegraphics[width=1.0\textwidth]{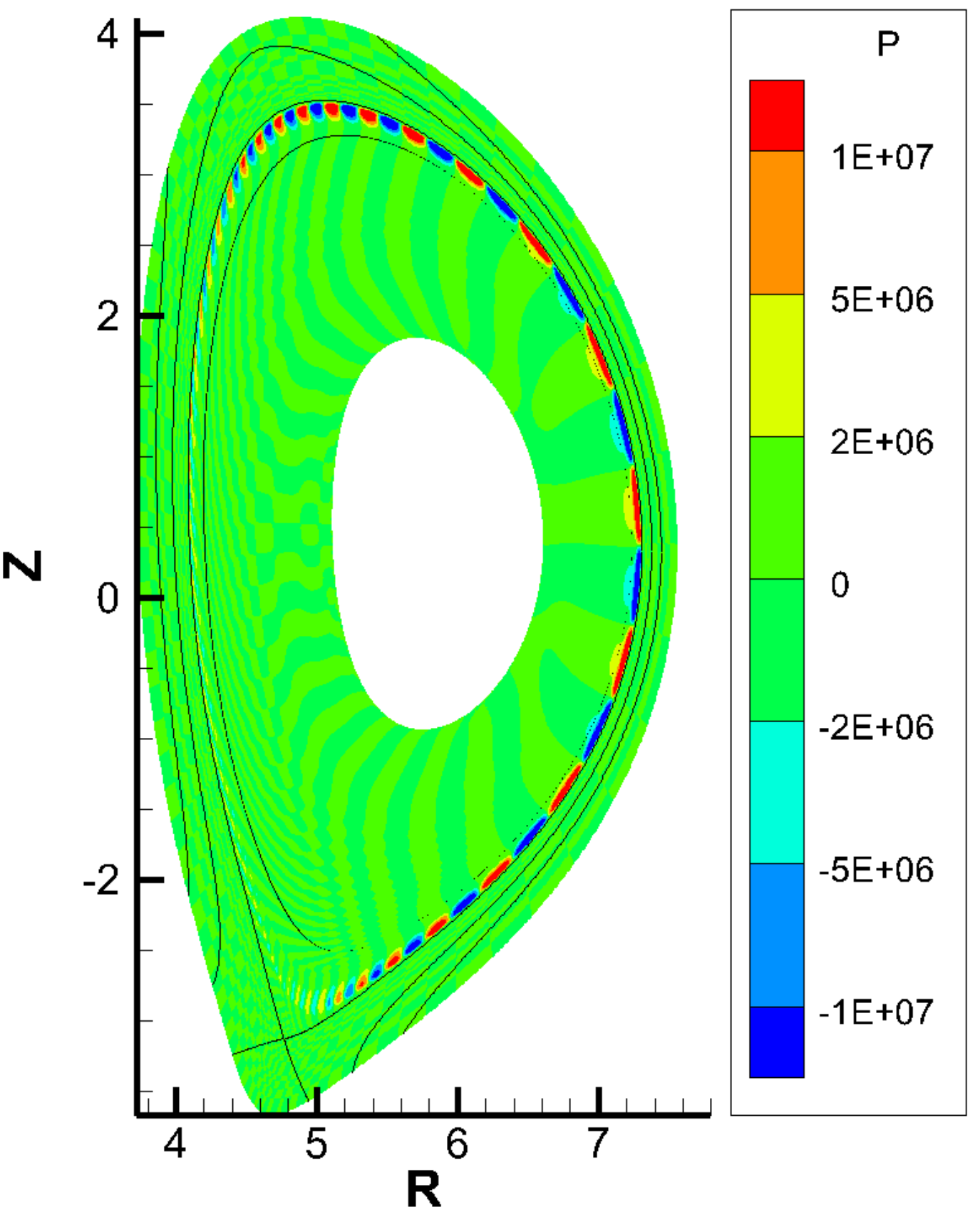}
\put(-220,240){\textbf{(d)}}
\end{minipage}

\caption{Dominant linear mode structures in presence of Spitzer resistivity profile and self-similar conducting wall at position $r_w=1.2a$ as shown in the color contours of: (a) perturbed $B_r$ of $n=3$ mode, and (b) perturbed pressure of $n=3$ mode, and (c) perturbed $B_r$ of $n=10$ mode, and (d) perturbed pressure of $n=10$ mode. The poloidal magnetic flux of the equilibrium is shown as the dark line contours in each plot.}
\label{fig_con_spitzer}
\end{figure}

All unstable modes have radial structure only localized at the edge pedestal region, close to the inside of separatrix which is indicated by black lines of poloidal flux contour (Figs.~\ref{fig_con_spitzer} - \ref{fig_con_realwall}). The positions and shapes of mode structure of these two different wall shapes are essentially the same. The spatial structure of the $n=10$ mode is more radially localized than that of the $n=3$ mode.

\begin{figure}[htbp]
\centering
\begin{minipage}{0.49\textwidth}
\includegraphics[width=1.0\textwidth]{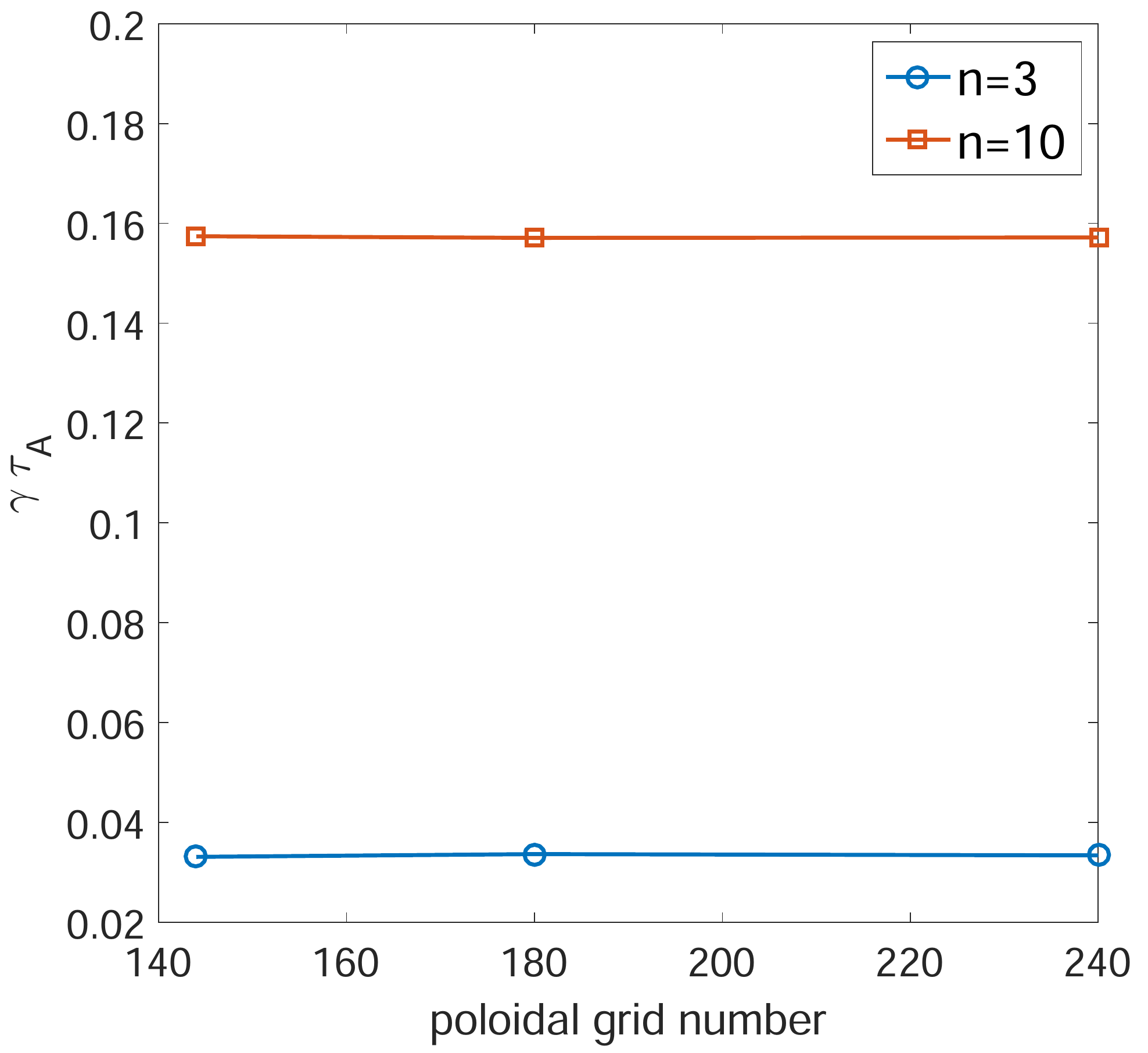}
\put(-220,190){\textbf{(a)}}
\end{minipage}
\begin{minipage}{0.49\textwidth}
\includegraphics[width=1.0\textwidth]{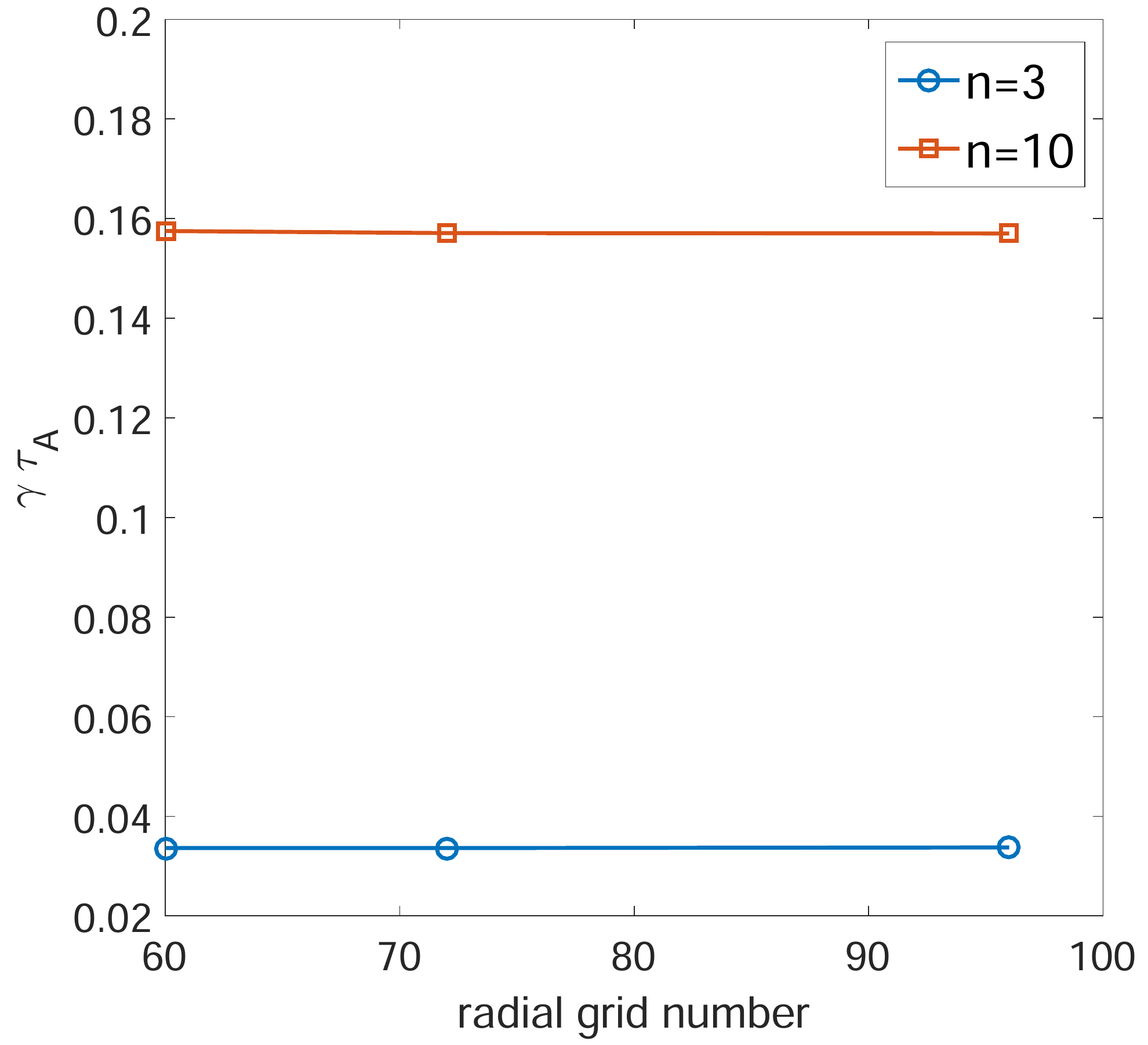}
\put(-220,190){\textbf{(b)}}
\end{minipage}

\begin{minipage}{0.49\textwidth}
\includegraphics[width=1.0\textwidth]{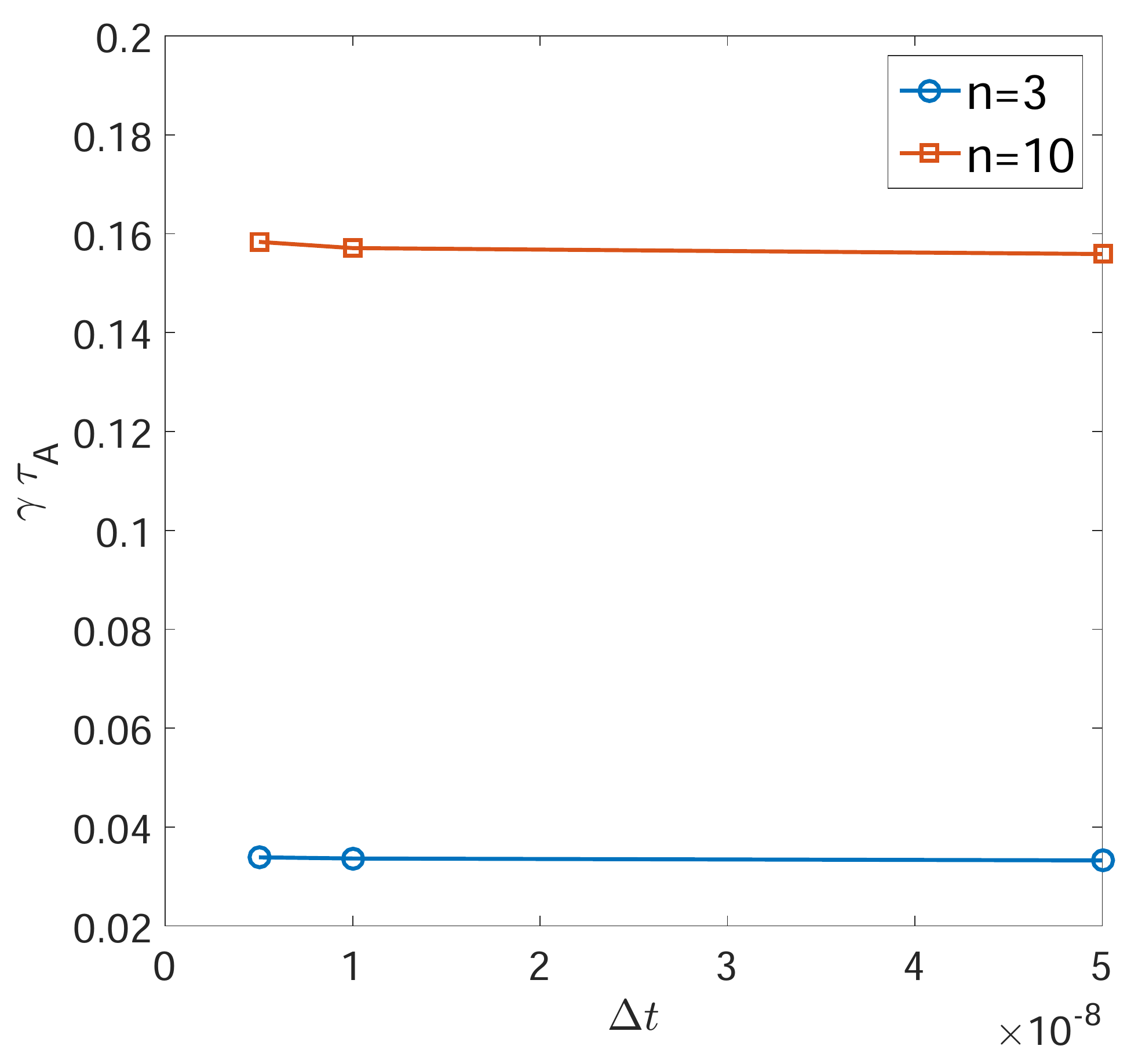}
\put(-220,190){\textbf{(c)}}
\end{minipage}
\begin{minipage}{0.49\textwidth}
\includegraphics[width=1.0\textwidth]{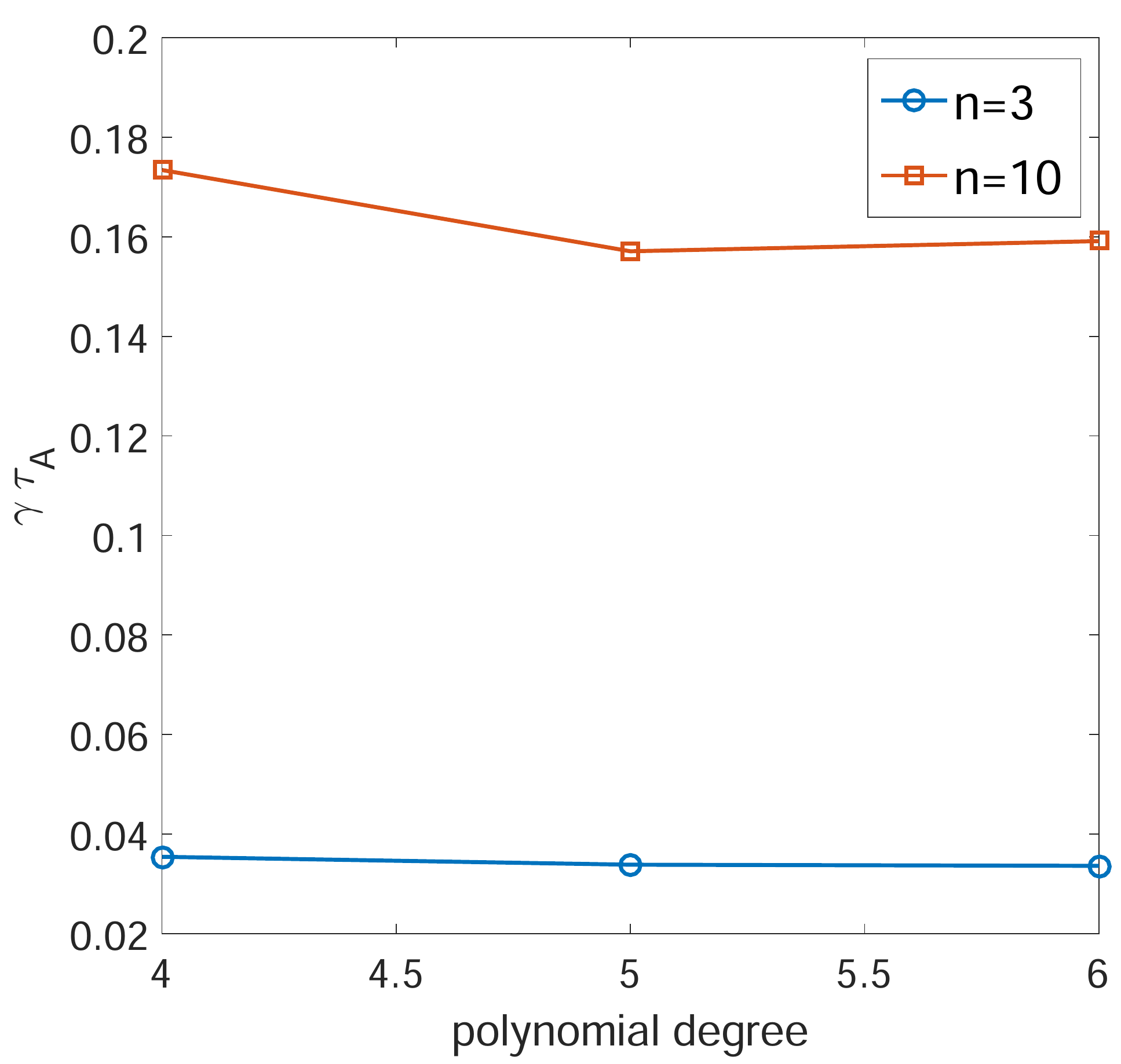}
\put(-220,190){\textbf{(d)}}
\end{minipage}

\caption{Linear growth rates as functions of (a) poloidal grid number, (b) radial grid number, (c) time step, and (d) polynomial degree for $n=3$, $10$ modes from NIMROD calculations.}
\label{fig_dtm}
\end{figure}

A thorough convergence has been checked for radial and poloidal grid numbers, time step ($\Delta t$) and polynomial degree of finite element basis used in NIMROD calculation. The growth rates of modes $n=3,10$ remain almost same for poloidal grid number range $150-240$ (Fig.~\ref{fig_dtm}a) and radial grid number range $60-96$ (Fig.~\ref{fig_dtm}b). From time step $\Delta t = 5 \times 10^{-9}{\rm s}$ to $\Delta t = 5 \times 10^{-8}{\rm s}$ (Fig.~\ref{fig_dtm}c) the variation in growth rate remains within $1\%$. Although there is moderate difference in growth between polynomial degree $4$ and $5$ for mode $n=10$, but polynomial degrees $5$ and $6$ have almost same growth rates (Fig.~\ref{fig_dtm}d). These results show a good numerical convergence in NIMROD calculation.


\section{Rotational stabilization on resistive wall mode}
\label{sec6}

The $\beta_N$ in baseline scenario is $1.88$, well below the no-wall $\beta$ limit expected from the experimental scaling law $\beta_{N,no-wall} \sim 4l_{\rm i}=2.52$, where $l_{\rm i}$ is the plasma inductance~\cite{strait1994}. However, the above results from both NIMROD and AEGIS suggest that at no-wall limit, the long-wavelength $n=1$ mode in CFETR baseline scenario could be unstable. This goes contrary to the expectation that normally such a low $\beta_N$ would help this equilibrium to lie within stability limits of global ideal MHD modes. The strong reversed shear in core region could be the main reason for the difference between the expection from the scaling law and our calculation results in terms of the no-wall $\beta$ limit~\cite{manickam1994}.

\begin{figure}[htbp]
\centering
\includegraphics[width=0.8\textwidth]{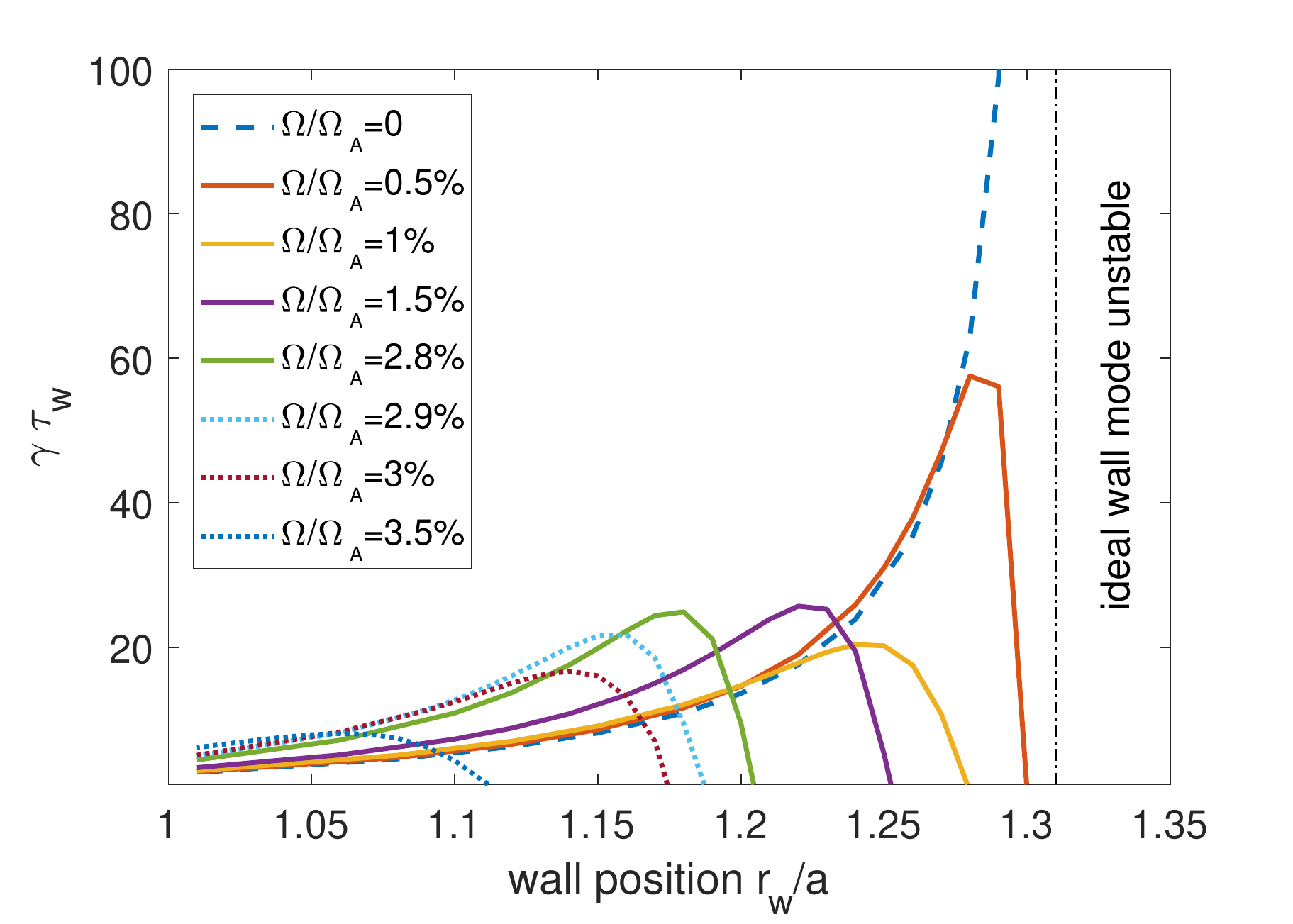}

\caption{The $n=1$ RWM growth rates as functions of wall position for different toroidal rotation frequencies. The critical wall position for ideal-wall external kink mode is plotted as the vertical dot-dashed line.}
\label{fig_rwm}
\end{figure}

The linear growth rate of $n=1$ RWM is calculated using AEGIS code. In the case of static equilibrium, the growth rate monotonically increases with the wall position from plasma boundary to $r_c=1.31a$ (the blue dashed line in Fig.~\ref{fig_rwm}). The resistive diffusion time of wall $\tau_w=\mu_0 \bar{r_w} d/\eta_w$ is used for normalizing the RWM growth rates, where $\bar{r_w}$ is the average minor radius of wall, $d$ is the wall thickness, and $\eta_w$ the wall resistivity. This $r_c$ is the critical wall position (vertical dashed line) where the mode turns into fast growing ideal-wall mode.

\begin{figure}[htbp]
\centering
\begin{minipage}{0.49\textwidth}
\includegraphics[width=1.0\textwidth]{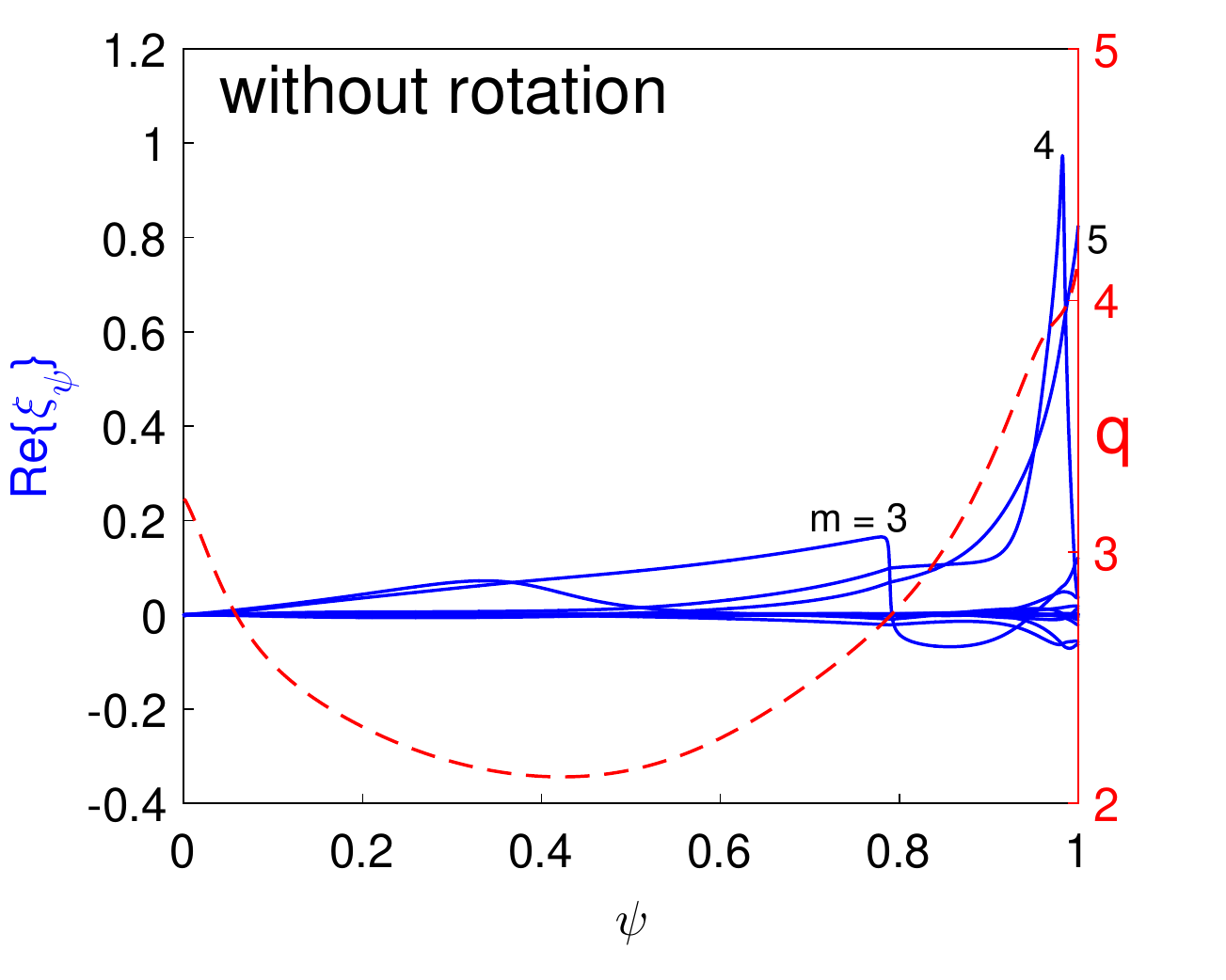}
\put(-220,150){\textbf{(a)}}
\end{minipage}
\begin{minipage}{0.49\textwidth}
\includegraphics[width=1.0\textwidth]{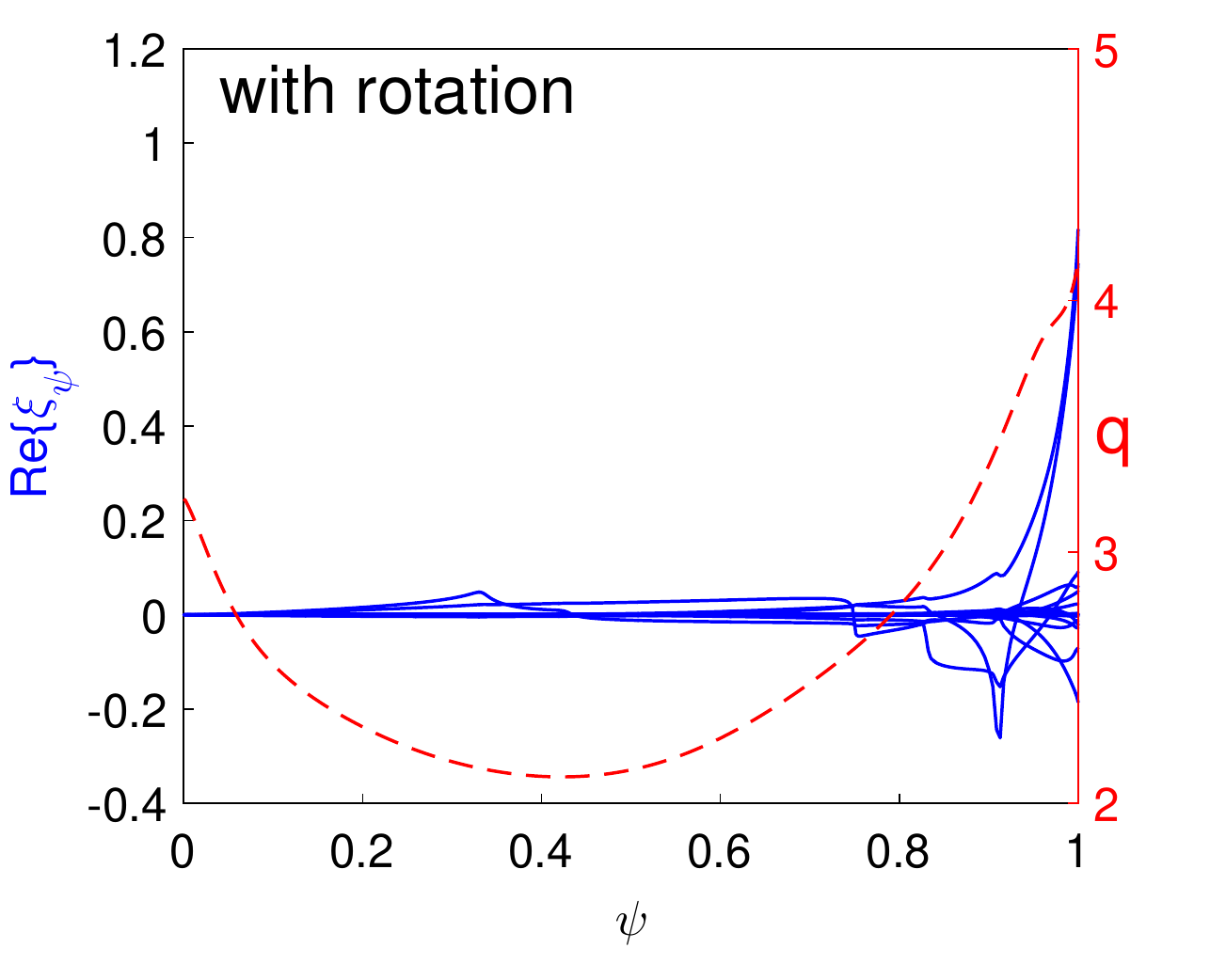}
\put(-220,150){\textbf{(b)}}
\end{minipage}

\caption{Real component of the radial displacements for the unstable $n=1$ RWM (a) in absence of rotation and (b) in presence of toroidal rotation (frequency $\Omega=2.9\% \Omega_A$) with the wall position $r_w = 1.2a$, respectively.}
\label{fig_eigenrwm}
\end{figure}

Long term steady-state operation requires stabilization of $n=1$ RWM. The toroidal rotation may open a stable window for RWM near the critical wall position. The width of the stable window is determined by rotation speed. Such results for CFETR are summarized in Fig.~\ref{fig_rwm} for rotation frequency from $\Omega = 0$ to $\Omega = 3.5\% \Omega_{\rm A}$, where $\Omega_{\rm A}$ is the Alfv\'en frequency evaluated at magnetic axis. For the designed wall configuration $r_w = 1.2a$, full stabilization can be achieved at $\Omega=2.9\% \Omega_{\rm A}$. The global mode structure becomes more localized to the edge region as $\Omega$ increases from $0$ to $2.9\% \Omega_{\rm A}$ (Fig.~\ref{fig_eigenrwm}). In the CFETR baseline scenario, the rotation frequency is around $15 - 80\, krad/s$ ($\sim 1\% - 5\%\Omega_{\rm A}$), sightly larger than the prediction for ITER ($10-20 \, krad/s$, $\sim 0.7\% - 1.5\%\Omega_{\rm A}$) in~\cite{jiale2017,chrystal17}.

\section{Summary and Discussions}
\label{sec7}

In summary, our linear stability analysis of CFETR baseline scenario using the initial-value code NIMROD and the eigen-value code AEGIS has found the dominant growth in each of the low-$n$ ($n=1-10$) modes with qualitative agreement between the two codes. The external kink component, which leads to the global mode structure in these low-$n$ modes, gradually reduces as $n$ increases. All the growth rates approach the corresponding no-wall limits when the wall moves sufficiently away. Effects of different plasma-vacuum profile models and wall shapes are examined. The Spitzer resistivity profile and the designed wall geometry are found more stabilizing, whereas the presence of an edge density pedestal tends to be destabilizing. 

For resistive wall, the $n = 1$ mode is found unstable for wall location $r_w<1.31a$. The toroidal rotation required for full suppression of the $n=1$ RWM with wall location $r_w=1.2a$ is determined to be $\Omega=2.9\%$. Effects of trapped and energetic particles~\cite{hu2004,hao2011}, as well as diamagnetic drift~\cite{zheng2017} may further reduce the rotation threshold for RWM stabilization. These additional passive stabilizing mechanisms are to be included next in stability analysis for CFETR baseline scenario.  

\ack

This work was supported by the Fundamental Research Funds for the Central Universities at Huazhong University of Science and Technology Grant No. 2019kfyXJJS193, the National Key Research and Development Program of China No. 2017YFE0300500, 2017YFE0300501, the National Natural Science Foundation of China Grant Nos. 11775221 and 51821005, U.S. DOE Grant Nos. DE-FG02-86ER53218 and DE-SC0018001. Author D.~B. was partially supported by CAS President International Fellowship Initiative (PIFI), the China Postdoctoral Science Foundation Grant No. 2016M592054 and the Anhui Provincial Natural Science Foundation Grant No. 1708085QA22. We are grateful for the support from the NIMROD team. This research used the computing resources from the Supercomputing Center of University of Science and Technology of China and the National Energy Research Scientific Computing Center, a DOE Office of Science User Facility supported by the Office of Science of the U.S. Department of Energy under Contract No. DE-AC02-05CH11231.


\section{References}
\providecommand{\newblock}{}

\newpage

\newpage
\begin{figure}[htbp]
\centering
\includegraphics[width=0.8\textwidth]{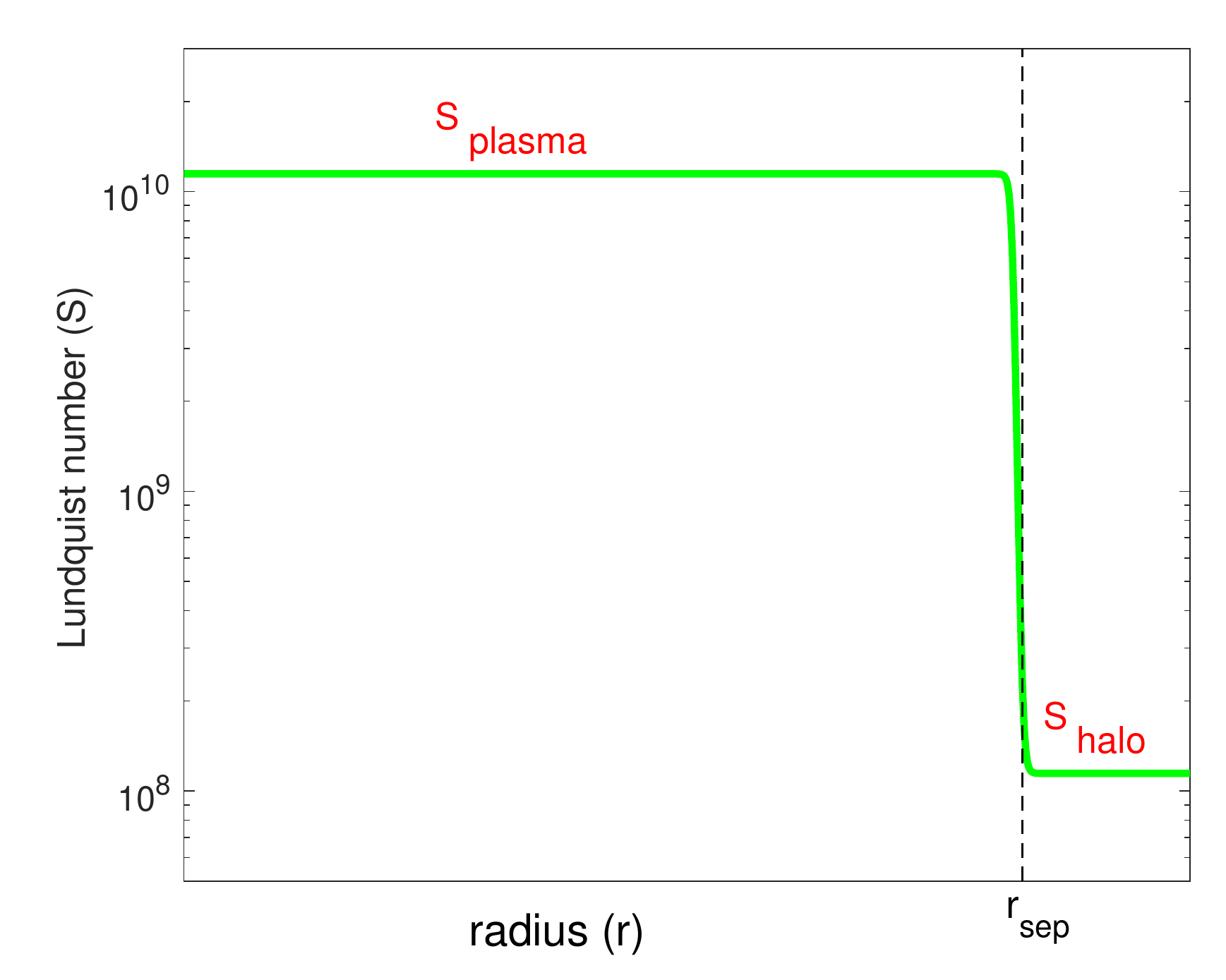}
\caption{The typical profile of Lundquist number used in NIMROD simulation. The value of $S_{\rm plasma}=1.146\times10^{10}$ and $S_{\rm halo}=1.146\times10^{8}$ are identified to model the ideal MHD limit of plasma and halo region. $r_{\rm sep}$ is the position of separatrix. }
\label{fig_S} 
\end{figure}

\newpage

\newpage

\newpage

\newpage

\newpage

\newpage

\newpage

\newpage

\newpage

\newpage

\end{document}